# Early Release Science of the exoplanet WASP-39b with JWST NIRCam

Eva-Maria Ahrer[1,2], Kevin B. Stevenson[3], Megan Mansfield[4,5], Sarah E. Moran[6], Jonathan Brande[7], Giuseppe Morello[8,9,10], Catriona A. Murray[11], Nikolay K. Nikolov[12], Dominique J. M. Petit dit de la Roche[13], Everett Schlawin[4], Peter J. Wheatley[1,2], Sebastian Zieba[14,15], Natasha E. Batalha[16], Mario Damiano[17], Jayesh M Goyal[18], Monika Lendl[13], Joshua D. Lothringer[19], Sagnick Mukherjee[20], Kazumasa Ohno[20], Natalie M. Batalha[20,21], Matthew P. Battley[2,1], Jacob L. Bean[22], Thomas G. Beatty[23], Björn Benneke[24], Zachory K. Berta-Thompson[11], Aarynn L. Carter[20], Patricio E. Cubillos[25,26], Tansu Daylan[27,28], Néstor Espinoza[12,29], Peter Gao[30], Neale P. Gibson[31], Samuel Gill[2], Joseph Harrington[32], Renyu Hu[33,34], Laura Kreidberg[14], Nikole K. Lewis[35], Michael R. Line[36], Mercedes López-Morales[37], Vivien Parmentier[38,39], Diana K. Powell[37,5], David K. Sing[40,29], Shang-Min Tsai[38], Hannah R Wakeford[41], Luis Welbanks[36,5], Munazza K. Alam[30], Lili Alderson[41], Natalie H. Allen[29,42], David R. Anderson[2,1], Joanna K. Barstow[43], Daniel Bayliss[2], Taylor J. Bell[44], Jasmina Blecic[45,46], Edward M. Bryant[47], Matthew R. Burleigh[48], Ludmila Carone[25], S.L. Casewell[48], Quentin Changeat[49,12,50], Katy L. Chubb[51], Ian J.M. Crossfield[7], Nicolas Crouzet[52], Leen Decin[53], Jean-Michel Désert[54], Adina D. Feinstein[22,42], Laura Flagg[35], Jonathan J. Fortney[20], John E. Gizis[55], Kevin Heng[56,2,57], Nicolas Iro[58], Eliza M.-R. Kempton[59], Sarah Kendrew[49], James Kirk[37,60,61], Heather A. Knutson[34], Thaddeus D. Komacek[59], Pierre-Olivier Lagage[62], Jérémy Leconte[63], Jacob Lustig-Yaeger[3], Ryan J. MacDonald[35,64,5], Luigi Mancini[65,66,14], E. M. May[3], N. J. Mayne[67], Yamila Miguel[15,68], Thomas Mikal-Evans[14], Karan Molaverdikhani[56,69,14], Enric Palle[8], Caroline Piaulet[24], Benjamin V. Rackham[70,71,72], Seth Redfield[73], Laura K. Rogers[74], Pierre-Alexis Roy[24], Zafar Rustamkulov[40], Evgenya L. Shkolnik[36], Kristin S. Sotzen[3,40], Jake Taylor[38,24], P. Tremblin[75], Gregory S. Tucker[76], Jake D. Turner[35,5], Miguel de Val-Borro[77], Olivia Venot[78], Xi Zhang[79]

*Corresponding authors' emails: eva-maria.ahrer@warwick.ac.uk, kevin.stevenson@jhuapl.edu
All author affiliations are listed at the end of the paper

**Measuring the metallicity and carbon-to-oxygen (C/O) ratio in exoplanet atmospheres is a fundamental step towards constraining the dominant chemical processes at work and, if in equilibrium, revealing planet formation histories. Transmission spectroscopy[e.g., 1,2] provides the necessary means by constraining the abundances of oxygen- and carbon-bearing species; however, this requires broad wavelength coverage, moderate spectral resolution, and high precision that, together, are not achievable with previous observatories. Now that JWST has commenced science operations, we are able to observe exoplanets at previously uncharted wavelengths and spectral resolutions. Here we report time-series observations of the transiting exoplanet WASP-39b using JWST's Near InfraRed Camera (NIRCam). The long-wavelength spectroscopic and short-wavelength photometric light curves span 2.0 – 4.0 μm, exhibit minimal**

**systematics, and reveal well-defined molecular absorption features in the planet's spectrum. Specifically, we detect gaseous $H_2O$ in the atmosphere and place an upper limit on the abundance of $CH_4$. The otherwise prominent $CO_2$ feature at 2.8 μm is largely masked by $H_2O$. The best-fit chemical equilibrium models favour an atmospheric metallicity of 1–100× solar (i.e., an enrichment of elements heavier than helium relative to the Sun) and a sub-stellar carbon-to-oxygen (C/O) ratio. The inferred high metallicity and low C/O ratio may indicate significant accretion of solid materials during planet formation[e.g., 3,4] or disequilibrium processes in the upper atmosphere[e.g., 5,6].**

*JWST* has demonstrated the necessary precision and wavelength coverage to make bulk characterization of hot exoplanet atmospheres routine[7]. The *JWST* Director's Discretionary Early Release Science (ERS) program provides the scientific community with observations of typical targets quickly enough to inform planning for the telescope's second cycle of scheduled observations. The primary goals of the Transiting Exoplanet Community ERS program (ERS-1366, led by N. M. Batalha, J. Bean, and K. B. Stevenson) are to demonstrate instrument capabilities, quickly build community experience, and seed initial discovery in transiting exoplanetary science[8,9]. The Panchromatic Transmission program observed a single exoplanet, WASP-39b, in transmission using four different instrument modes. It included overlapping wavelength coverage to cross-compare and validate all three near-infrared (NIR) instruments for time-series observations. The observations presented here form one quarter of this program, demonstrating the capacity of the Near-Infrared Camera (NIRCam) for transiting exoplanet atmospheric characterisation.

WASP-39b is a highly inflated exoplanet of roughly Saturn mass, orbiting its G7 main-sequence star with a 4.05 day period[10]. We selected WASP-39b for its inactive host star and prominent spectroscopic features, which trace the atmospheric composition of the planet. We confirmed the star's relative inactivity through a photometric monitoring campaign using the Next-Generation Transit Survey (NGTS)[11] and Transiting Exoplanet Survey Satellite (TESS)[12] (see Methods). Reported atmospheric metallicities span a range of possible values (0.003 – 300× solar)[13–18] due to limits on wavelength coverage, lower signal-to-noise ratio data, and/or differences between analyses[19–22]. If the Solar System trend for gas giants[23,24] also applies to exoplanets, WASP-39b should have an atmospheric metallicity comparable to that of Saturn (10 × solar[25]) and other Saturn-mass exoplanets.

We observed a single transit of WASP-39b with *JWST*'s NIRCam instrument on 22-23 July 2022 (19:28 – 03:40 UT). The Grism R and F322W2 filter in the long-wavelength (LW) channel dispersed light from 2.420 – 4.025 μm at a spectral resolution *R* of 1570 - 2594 over 1023 resolution elements. The short-wavelength (SW) channel allowed the simultaneous measurement of light, i.e. photometry, spanning 2.0 – 2.2 μm using the WLP8 weak lens and F210M filter. See the Methods section for more details.

The team conducted three independent reductions of the NIRCam LW spectroscopic data and four independent fits and analyses of the reduced data. We also performed two independent analyses of the SW photometric data. For both data reductions (LW and SW), customising the *JWST* Science Calibration Pipeline (jwst) to allow for minor adaptations to default steps and values worked best (see Methods). The wavelength solution available with the reference files provided by the *JWST* Calibration Reference Data System (CRDS) at the time of our analysis was inaccurate (particularly for the blue edge of the LW channel), so we redefined our wavelength values using a polynomial wavelength calibration derived from a planetary nebula observed as part of commissioning (Program 1076).

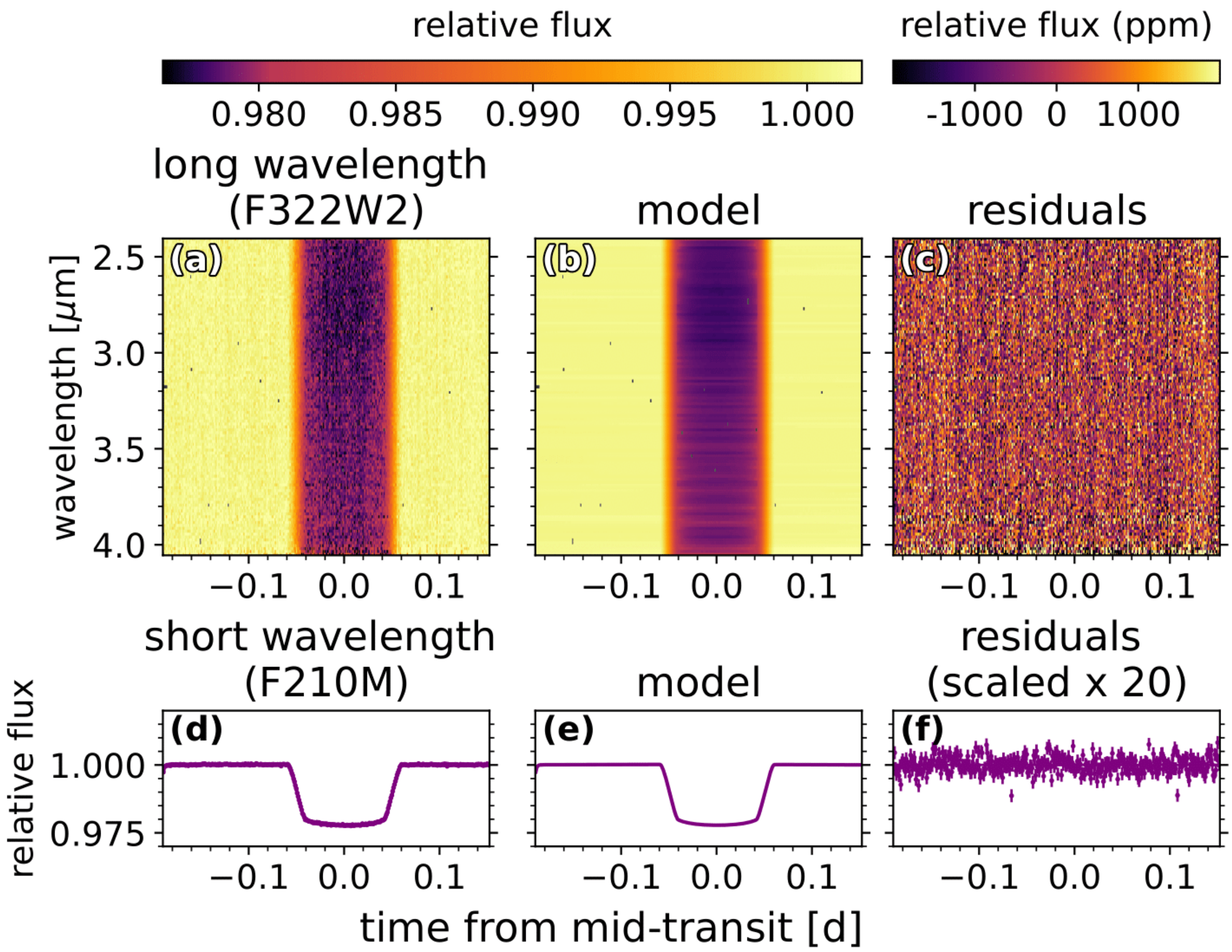

**Figure 1: The relative brightness of the WASP-39 planetary system as a function of time and wavelength, as measured by NIRCam.** Panels a, b, and c show the spectroscopic data while panels d, e, and f present the photometric (short wavelength) channel. The extracted flux normalised by the median stellar spectrum is shown in panels a and d, the best-fit transit and systematic models are shown in panels b and e, and the residuals are shown in panels c and f. The flux decrease results from the transit of exoplanet WASP-39b in front of its star. The subtle variation in transit depth around 2.8 μm is due primarily to water vapour in the planet's atmosphere. The vertical striping in the residuals is due to $1/f$ noise.

We found no large systematic structures affecting the LW light curves and a minuscule ramp at the start of the SW light curve (Figure 1, panel d). The only other systematic identified was $1/f$ noise (or pink noise, where $f$ is frequency), which describes the detector's correlated read noise[27]. For NIRCam, this manifests as weak structures in the dispersion direction, as shown in Figure 1c. We did not correct for $1/f$ noise in the final LW reduction because it did not impact the precision reached by individual spectroscopic light curves (compare `tshirt` and `Eureka!` in Figure 2 for analyses with and without $1/f$ noise corrections). We removed structures due to $1/f$ noise in the SW reduction (see Methods). We found a linear model in time was sufficient to detrend the data, which produced uncertainties 1.18× the photon noise limit (median of 135 ppm for the transit depths) at a binned spectral resolution of 15 nm (~15 pixels). Similarly, the photometric transit-depth precision was 1.35× the noise limit at 53 ppm. The residuals are Gaussian (see Extended Data Figure 5 in Methods).

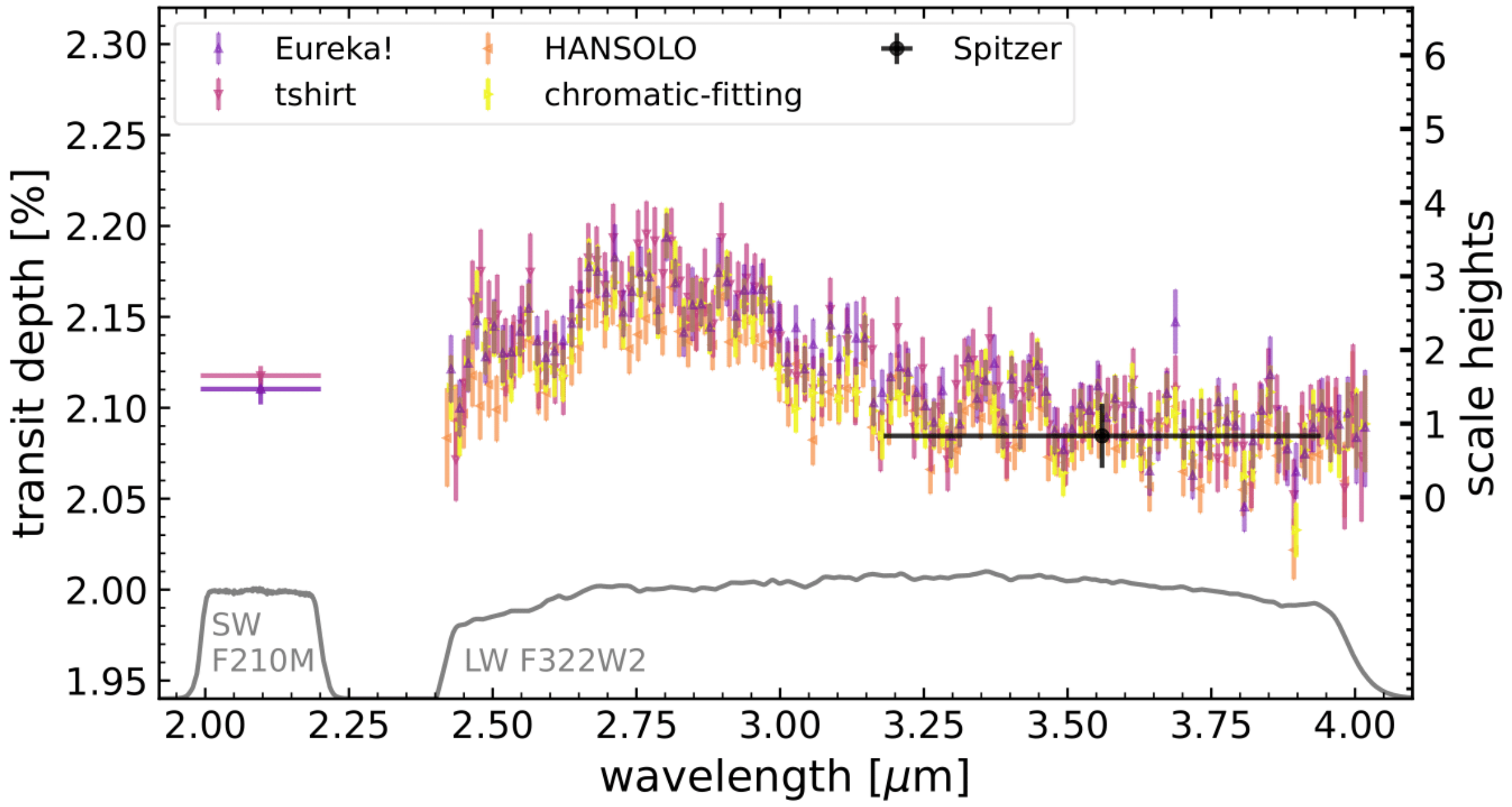

**Figure 2: The transit spectrum of WASP-39b as measured from JWST's NIRCam instrument.** The coloured points with 1σ uncertainties depict our independent analyses of the spectroscopic LW channel (2.420 – 4.025 μm) and photometric SW channel (2.0 – 2.2 μm) with their respective throughputs shown in grey. All analyses agree with the broadband Spitzer point (black circle, 3.2 – 4.0 μm). The broad feature centred at 2.8 μm spans 2.5 scale heights (~2,000 km) and is due primarily to water vapour within WASP-39b's atmosphere. We note the consistency between analyses in the fine structure.

Figure 2 displays the independently derived transit spectra and photometry. Each reduction is consistent with our selected reduction (`Eureka!`) to better than 1σ, as is the broadband 3.6 μm Spitzer point[13]. The overall shape of the spectrum is due primarily to absorption of water vapour (feature centred at 2.8 μm). The right-axis scale is in equivalent scale heights, where one scale height is approximately 800 km.

To interpret the presence of other molecules within the planetary atmosphere, we compared the `Eureka!` transit spectrum to a set of independently computed atmospheric model grids that spanned a range of cloud properties, metallicity values, and C/O ratios (see Methods).

Figure 3 shows a representative best-fit model highlighting the contributions of major molecular absorbers.

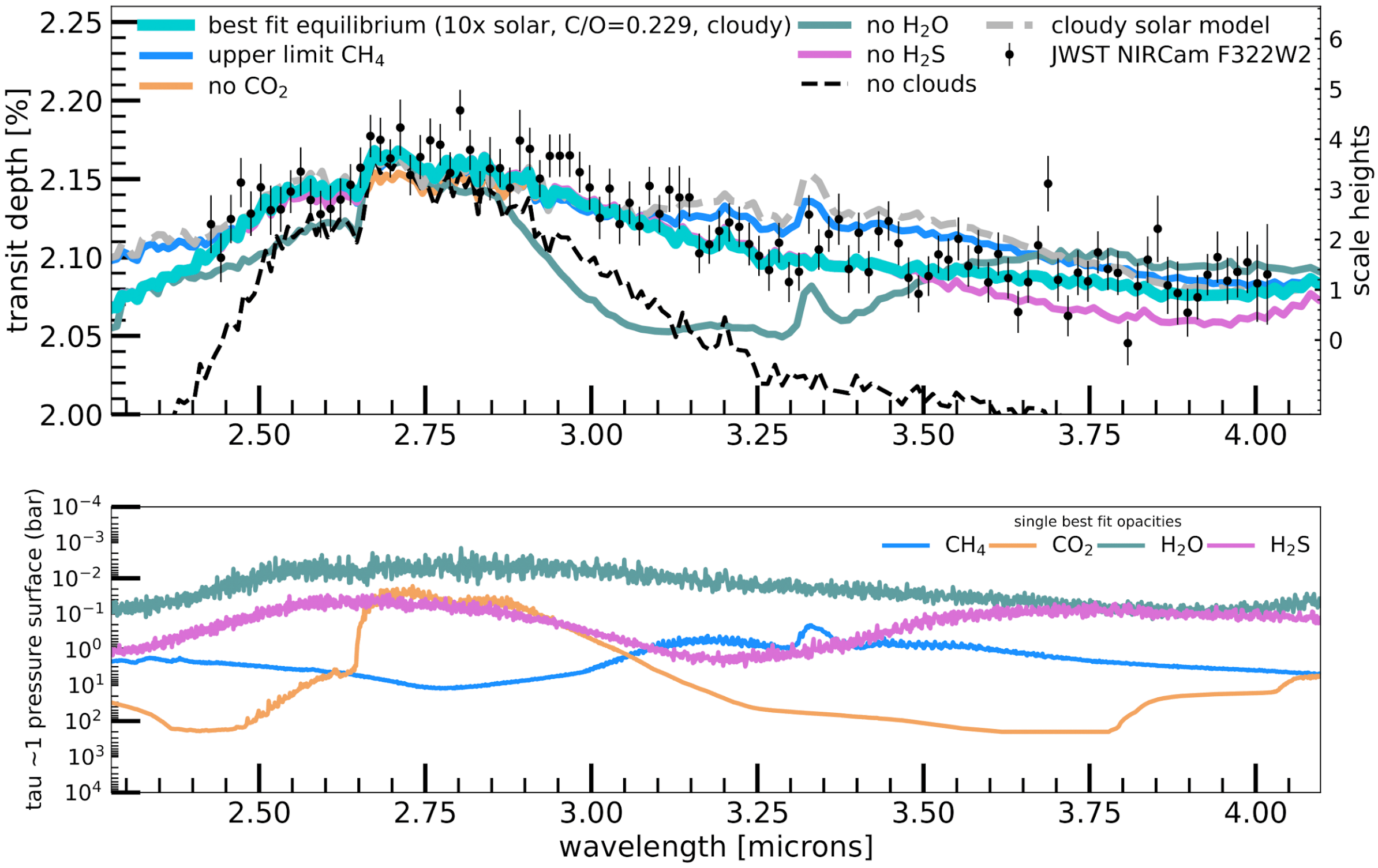

**Figure 3: Contributions of key absorbers impacting the spectrum.** Top: The best-fit `PICASO 3.0` equilibrium model (10× solar, C/O = 0.229, moderate grey clouds with cloud optical depth of $2.5\times10^{-3}$) is shown compared to the `Eureka!` reduction, along with models with individual molecular species removed to show its contribution to the spectrum. Each model is normalized to the data for illustration by offsetting each model to have the same transit depth at 2.8 µm. Water predominately sets the shape of the spectrum, followed by the influence of clouds. The grey dashed line shows a cloudy solar-metallicity and stellar-C/O atmospheric model, illustrating the lack of a strong $CH_4$ peak seen in the data. Bottom: The opacities of the dominant molecular species at an optical depth of 1 in the atmosphere. In the single best-fit model shown in the lower panel, the methane peak at 3.3 µm is blended out by water absorption. However, manual scaling of $CH_4$ gives an upper limit of $CH_4$ abundance (blue line) for the single best-fit model shown in the top panel.

Our spectroscopic wavelength range covered by NIRCam/F322W2 includes absorption features due to prominent atmospheric molecules such as $H_2O$, $CO_2$, and $CH_4$. From our model grid search, we definitively confirm the presence of $H_2O$ at nearly 16σ. Water vapour was previously identified in the atmosphere of WASP-39b using Hubble Space Telescope (*HST)* WFC3 observations taken at shorter wavelengths ($[H_2O] = -1.37^{+0.05}_{-0.13}$)[13]. We also see weak evidence for $CO_2$ absorption, previously seen with high confidence using

NIRSpec/PRISM at 4.3 μm[7], but the overlap between the $CO_2$ feature at 2.8 μm and the broad $H_2O$ feature (illustrated in Figure 3) leads to a more tentative identification here. Each forward model grid prefers significant cloud coverage, which impacts the spectrum at ~mbar pressures, despite differing cloud parametrisations between grids with varying levels of physical complexity (see Methods).

In a hot (~1000 K) solar-metallicity atmosphere with a stellar C/O ratio, $CH_4$ would be visible as a strong peak at 3.3 μm (grey dashed line in Figure 3, Extended Data Figure 7 in Methods) under thermochemical equilibrium. Such a peak is absent in the reduced spectrum. We quantified this using a residual fitting test (see Methods). In a higher-metallicity and/or lower-C/O atmosphere, carbon is increasingly partitioned into CO and $CO_2$, and the $CH_4$ peak at 3.3 μm disappears. Therefore, the absence of a strong $CH_4$ peak at 3.3 μm in our data drives the metallicity to higher values and the C/O ratio to lower values. We scaled the $CH_4$ volume mixing ratio (VMR) within our single best-fit `PICASO 3.0` model (10× solar metallicity; C/O ratio of 0.229) to determine an upper limit on the abundance of $CH_4$ at 1 mbar, where it contributes most strongly to the spectrum. Within our single best-fit model scaling, we find an upper limit on $CH_4$ abundance at 1 mbar of $5.5\times10^{-5}$ (or 55 ppm) VMR, above which the goodness of fit per free parameter, $\chi_\nu^2$, gets increasingly worse (i.e., $\chi_\nu^2 > 2$). We also tested whether other data reductions favoured best-fit models with stronger methane abundances, but found they did not have any statistical significance.

Driven by this $CH_4$ upper limit, the single best fit from each grid favours the lowest C/O ratio (0.229, 0.3, and 0.35 for `PICASO 3.0`, `PHOENIX`, and `ATMO`, respectively) within that grid. These best-fit point values for C/O from the three grids agree well with the value of $0.31^{+0.08}_{-0.05}$ found by Ref. [13]. We examined the effect of an even lower C/O grid point by computing the best-fit `PICASO 3.0` model with a C/O of 0.115, but found no discernible difference in the transit spectrum. Comparing our inferred C/O ratio for WASP-39b's atmosphere to that of its host star, we see that it is substellar (≤ 0.35, whereas WASP-39 is 0.46±0.09,[23]). We also note that the C/O ratio shown here represents the carbon-to-oxygen fraction of the planet's upper atmosphere rather than that of the whole atmosphere, as these NIRCam observations probe approximately the 0.1 – 10 mbar pressure range. WASP-39b's temperature-pressure profile is cool enough for the formation of silicate (i.e., O-bearing) cloud species at depth, which would deplete oxygen from the upper atmosphere and actually *increase* the C/O ratio aloft compared to the bulk planetary envelope[28,29].

Figure 4 compares our best-fit metallicity values, shown as separate O and C abundances, and C/O ratios to previous studies using *HST* data, as well as results for exoplanets observed at high resolution and Solar System gas giants. The *JWST*/NIRCam data rule out a super-stellar C/O ratio for WASP-39b. Additionally, Figure 4 demonstrates *JWST*'s capability to measure the C/O ratios of giant planet atmospheres by observing both O- and C-bearing species , which until now has only been achieved through high-resolution exoplanet observations[e.g., 30,31]. Similar measurements have been difficult to achieve from *HST* alone. Even in the Solar System gas giants, such constraints have proved difficult from both remote sensing and *in-situ* missions, as the low temperatures of Jupiter, Saturn, Uranus, and Neptune lead to condensation of most oxygen-bearing species (e.g., $H_2O$, $CO_2$) at high altitudes, prohibiting accurate measurement of the O abundance[e.g., 32,33].

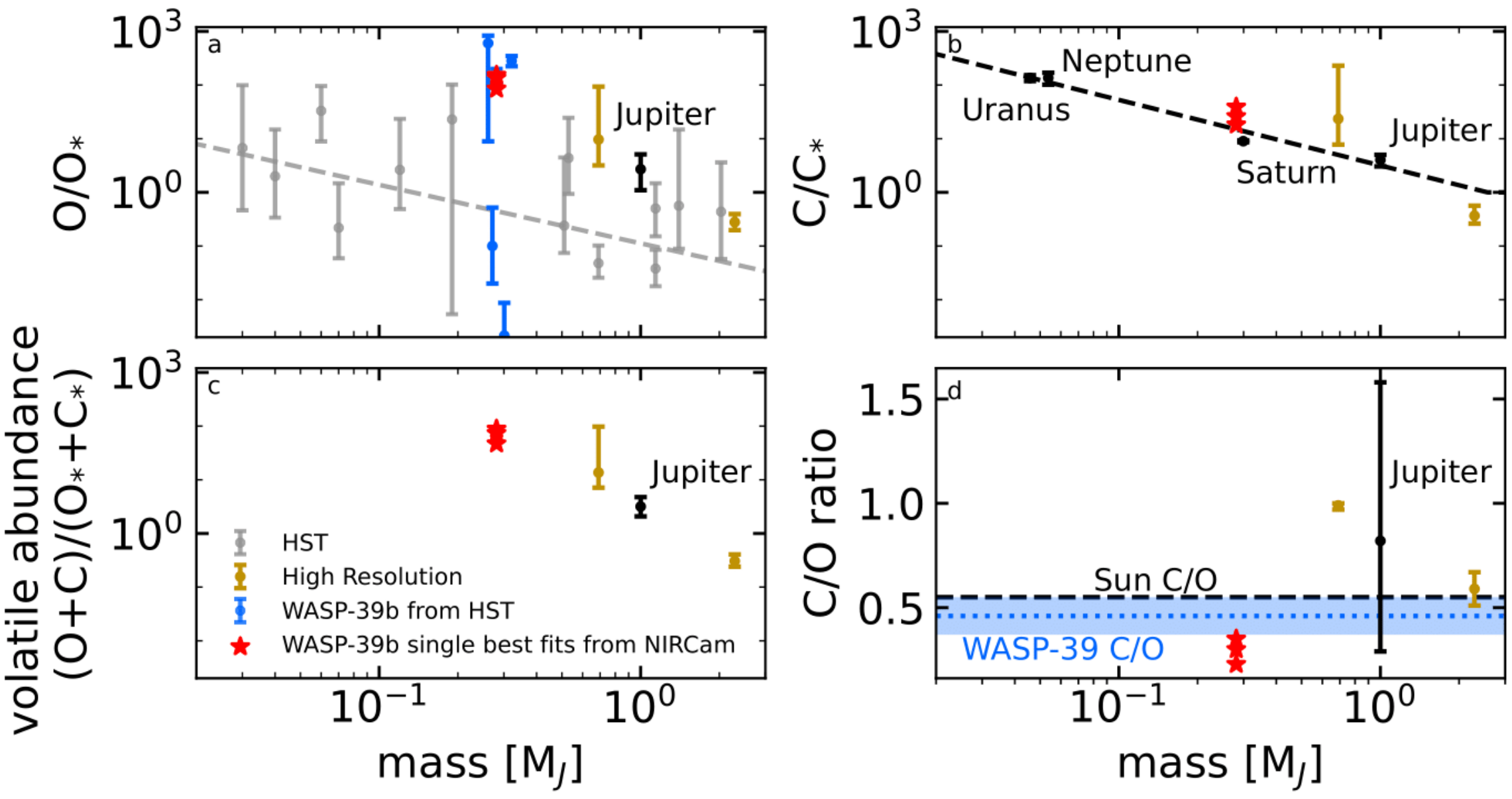

**Figure 4: Trends in elemental abundances and C/O ratio with planet mass.** Panels a, b, and c show the abundances of oxygen, carbon, and net volatiles (oxygen + carbon), respectively, scaled to stellar values. Grey points in panel (a) show HST constraints based on ≥2σ $H_2O$ detections, with the grey dashed line showing the best-fit trend from Ref. [18]. Blue points show all previous estimates of the metallicity of WASP-39b from HST data, offset in mass for clarity[13,15–18]. The black points and dashed line in panel b show a fit based on $CH_4$ abundances of Solar System giant planets[26,41–43]. Of the Solar System planets, only Jupiter has a constrained oxygen abundance (from Juno observations of $H_2O$,[33]). Gold points indicate high-resolution observations of $H_2O$ and CO in exoplanets[30,31], and red stars show the best-fit values for WASP-39b as measured by JWST/NIRCam for each of the three model grids described in this paper. The black dashed line in panel d depicts the solar C/O ratio of 0.55[44] and the blue dotted line with a shaded 1σ uncertainty region indicates the measured C/O ratio of the star WASP-39[23]. Our results for WASP-39b favour a super-stellar volatile abundance and substellar C/O ratio. However, we emphasize that a full retrieval will be necessary to determine accurate means and 1σ error bars for the NIRCam results.

The apparent substellar C/O ratio inferred from chemical equilibrium models may trace photochemical processes in the planet's upper atmosphere. For example, photochemical destruction of $CH_4$ in the upper atmosphere could explain the absence of a $CH_4$ peak at 3.3 μm[e.g., 6,34]. The most likely immediate products of $CH_4$ photolysis, such as HCN or $C_2H_2$, would be produced in abundances too small (≲ a few parts-per-million (ppm),[6,34]) to be robustly detected with a single NIRCam transit, even from complete $CH_4$ conversion. Alternatively, much of the carbon available from $CH_4$ photolysis could have been oxidized by photodissociated $H_2O$ to form CO and $CO_2$[6,34–36], though the absolute abundances of these two carbon reservoirs would not have been meaningfully altered since their abundances under chemical equilibrium are already higher than that of $CH_4$. Other proposed disequilibrium chemistry processes could reduce the $CH_4$ abundance at the terminator without also decreasing the C/O ratio[5,37–40]. We defer the exploration of complex disequilibrium models to atmospheric retrieval analyses using the full set of data provided by the Transiting Exoplanet Community ERS program. That dataset will also constrain the presence of additional oxygen- and carbon-bearing species to provide a more robust constraint on the C/O ratio than we can obtain here. However, the C/O ratio estimate we report from NIRCam is broadly consistent with the C/O ratio found from the other individual ERS WASP-39b datasets, which range from best-fits that are sub-solar (NIRISS/SOSS, Feinstein et al., submitted; NIRSpec/PRISM 3.0 - 5.0 μm,[7]; NIRSpec/G395, Alderson et al., submitted) to a slightly super-solar upper limit (NIRSpec/PRISM 0.5 - 5.5 μm, Rustamkulov et al., submitted).

If disequilibrium chemistry is not prevalent in the planet's upper atmosphere, the inferred high metallicity and low C/O ratio can be tied back to WASP-39b's formation. The most prominent scenario is that WASP-39b formed via core accretion exterior to the water-ice line and accreted low-C/O solid material in situ and/or while migrating inward within the protoplanetary disk[4,45,46]. Taken as such, JWST observations could offer important clues regarding the degree to which hot Jupiter atmospheres undergo solid accretion during their early evolution.

Here, we have demonstrated the excellent performance of NIRCam for exoplanet transmission spectroscopy. With the first *JWST* exoplanet spectra now comparable to the first near-infrared Jupiter spectra[47], the future promises many exciting discoveries and major advancements in the formation, evolution, and atmospheric chemistry of hot Jupiters.

# Methods

As part of this article's Reproducible Research Compendium, located on Zenodo at https://doi.10.5281/zenodo.7101283, we provide a Jupyter Notebook with step-by-step data reduction instructions replicating our chosen analysis, saved outputs from various pipeline stages, and the data used to generate relevant figures.

**Photometric Monitoring of Host Star**

In order to confirm that WASP-39 is a relatively inactive star, and that the *JWST* observations were not adversely affected by stellar activity, we carried out photometric monitoring with the ground-based Next Generation Transit Survey (NGTS)[11]. Monitoring began at the end of April 2022 and continued until late August, spanning the *JWST* ERS transit observations of WASP-39b in July. We used one camera on most photometric nights to take a series of 10 s images lasting on average for 2 h. The resulting monitoring light curve is plotted in Extended Data Figure 1 (top panel), showing one binned point for each night. Also included is the *TESS* sector 51 PDCSAP (Pre-search Data Conditioned Simple Aperture Photometry) light curve of WASP-39[12], which is binned to 2 h to be comparable with NGTS. Both light curves have been detrended against sky brightness. They show evidence for stellar activity, but only with a low amplitude of 0.06% in NGTS.

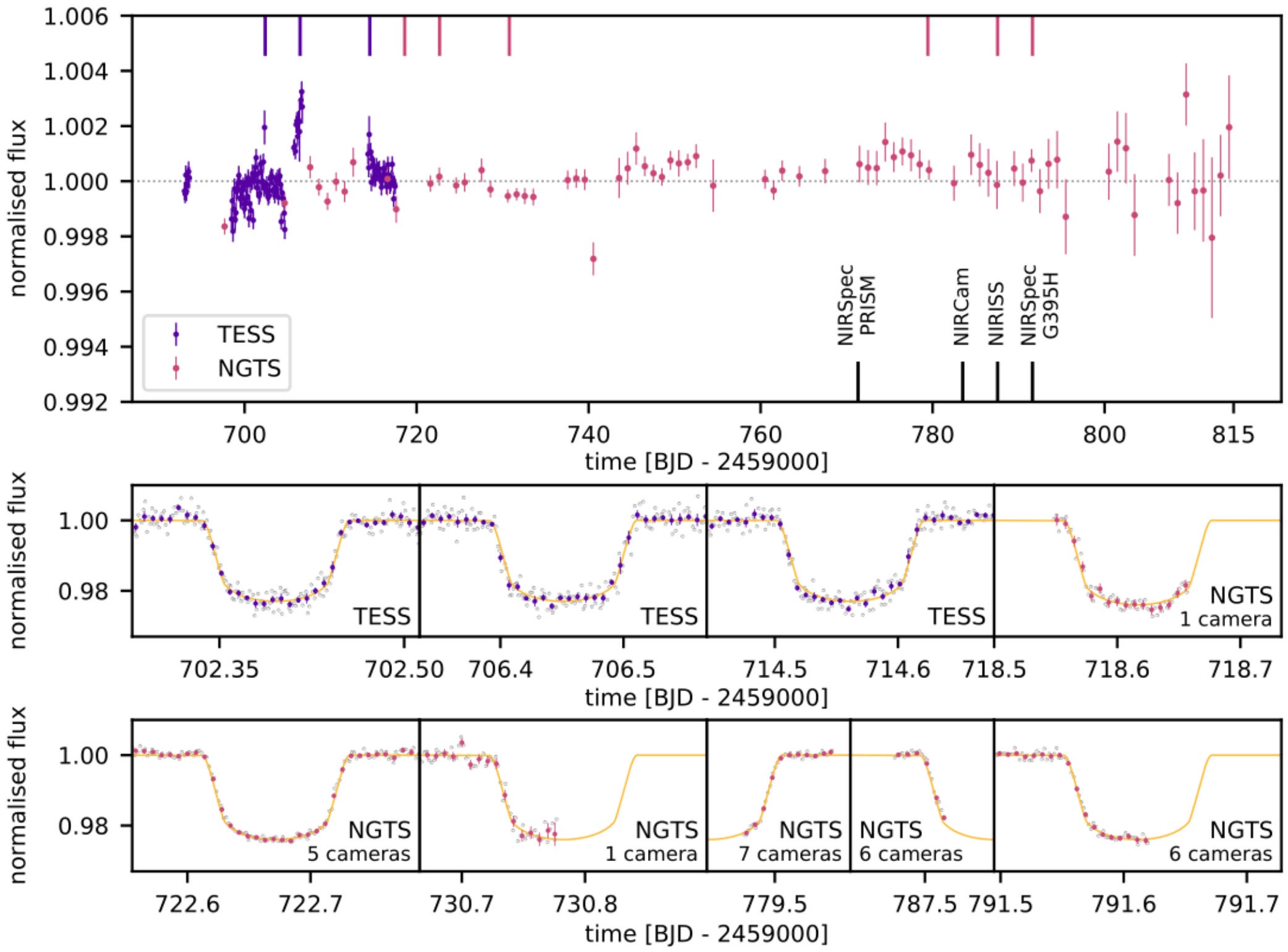

**Extended Data Figure 1: Photometric monitoring of WASP-39 (top) and individual transit observations (bottom) using NGTS (magenta) and TESS (dark purple)**. The black marks indicate the times of the four JWST ERS transit observations. The monitoring light curve shows evidence for optical variability, but with an RMS amplitude of only 0.06%

in NGTS. The times of the individual transit observations are indicated on the top panel, and they are all consistent with transits free of starspot crossings or other features associated with stellar activity.

Also plotted in Extended Data Figure 1 (lower panels) are individual transit observations of WASP-39b with NGTS and *TESS* (the times of which are indicated on the monitoring light curve). For four of the NGTS transits, we employed multiple cameras. This significantly improves the photometric precision[48], which is otherwise limited by atmospheric scintillation[49]. The transit models were generated from the system parameters listed in Extended Data Table 1. We only fit the transit times and the mutual depth of the *TESS* transits, which is slightly shallower than expected.

The transit observations in Extended Data Figure 1 show no evidence for starspot-crossing events, which would be visible as bumps in the transit light curve. The absence of such events across multiple high-precision transits provides additional evidence that WASP-39 is a quiet star and that the *JWST* ERS transit observations are unlikely to be adversely affected by stellar variability.

**_JWST_ NIRCam Observation**

*JWST* observed the 2.8-hour transit of WASP-39b over a span of 8.2 hours, providing a baseline before and after transit to measure transit depths accurately. A dichroic beam splitter allows NIRCam to simultaneously observe a target in both short wavelength (SW) and long wavelength (LW) channels[50,51]. The LW channel used the Grism R + F322W2 filter to observe a wavelength range of 2.420 – 4.025 μm with a spectroscopic resolving power of $R{\sim}1,600$ at 4 μm (Extended Data Figure 2, top panel). The SW imaging channel used the WLP8 weak lens and F210M filter (2.0 – 2.2 μm) to produce the hexagonal pattern shown in the bottom panel of Extended Data Figure 2. Spreading the light prevents saturation, reduces variability due to image motion over an imperfect flat field, and allows monitoring of mirror-segment alignment. Both SW and LW channels used the SUBGRISM256 subarray mode with four output amplifiers and the SHALLOW4 readout pattern to minimize data volume. With 12 groups per integration (82.17 s total), we acquired 366 integrations for this transit observation.

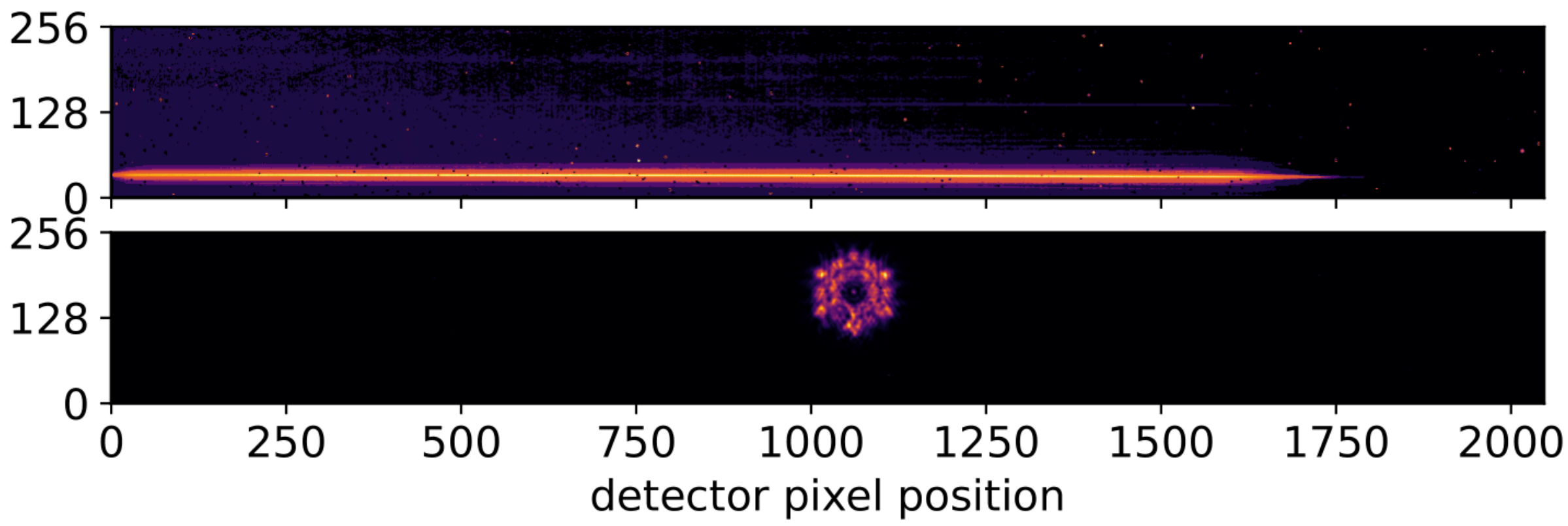

**Extended Data Figure 2: Raw NIRCam image of the LW (top) and SW (bottom) channels.** The faint horizontal stripes seen in the LW channel originate from neighbouring objects. The SW channel is able to track changes in alignment for individual mirror segments. No impactful tilt events were noted in this observation.

**Data Reduction and Calibration**

We conducted independent data analyses using multiple pipelines and fitting tools to ensure that we obtained the same transmission spectrum using different reduction pipelines. We also varied the fitting methods within a given data reduction pipeline.

Many of the reductions presented below used intermediate data products from or made minor edits to the *JWST* Science Calibration Pipeline (jwst)[1], which we briefly summarize here. jwst is a Python software suite for processing data from all *JWST* instruments and observing modes, and is organised into three stages. Stage 1 takes in uncal.fits files and performs detector-level corrections and ramp fitting for individual exposures (i.e., ramps-to-slopes conversion; these ramps are the flux increases during an exposure, not to be confused with baseline ramps over the course of the entire transit). Stage 2 takes in slope images (ramps) from Stage 1 and performs an assignment of the world coordinate system, flat fielding, and assignment of a wavelength solution. Stage 3 takes in calibrated 2D images from Stage 2 and extracts a time series of 1D spectra. The default pipeline settings include a flux calibration step at Stage 2. In all data reductions presented below, we skipped that step, as it introduced scatter in the extracted spectral time series. This is justified because the transit depths we compute are relative, rather than absolute, flux measurements.

Below we describe the independent data reductions applied to the SW photometry and LW spectroscopy, respectively. In each case we note where data reductions deviated from the standard jwst pipeline.

[1] https://jwst-pipeline.readthedocs.io/

**Short-Wavelength Photometry**
We performed two independent short wavelength data reductions using the open-source `Eureka!` and `tshirt` pipelines.

### Eureka! Short Wavelength Reduction

`Eureka!` is an open-source pipeline designed to perform spectral extraction and fitting for *JWST* exoplanet time-series observations[52]. The `Eureka!` short wavelength data reduction used the default jwst settings for Stages 1 and 2, with the exception of increasing the rejection threshold during jump detection to 10σ, which improved the quality of the resulting light curve.

In Stage 3, we first masked all pixels for which the "DO_NOT_USE" data quality flag was raised by the jwst pipeline. We then performed an outlier rejection along the time axis for each individual pixel in a segment using a 7σ threshold, repeating this process twice. Next, we corrected for the $1/f$ noise in each of the four amplifier regions by subtracting the median flux in each row calculated without pixels containing the star. We interpolated over flagged pixels using a cubic function. Finally, we determined the image centre and performed aperture photometry on the target. We explored different target apertures and background annuli, and chose the combination that minimised the root-mean-square variations, leading to a target aperture radius of 65 pixels and a background annulus from 70 to 90 pixels relative to the centre.

### tshirt Short-Wavelength Reduction

`tshirt` is an open source pipeline[2] that has tools to modify jwst and performs photometric and optimal spectral extraction of light curves.

In the Stage 1 SW analysis, `tshirt` applied a row-by-row, odd/even-by-amplifier (ROEBA) subtraction algorithm that used background pixels to reduce the $1/f$ noise. In this procedure, background pixels are used to correct each group in a similar fashion as with reference pixel correction.[3] The ROEBA correction happens after the bias subtraction step. First, the median of all even columns' background rates is subtracted from all even columns and the median of all odd columns' background rates is subtracted from all odd columns to remove most pre-amp reset offsets and odd/even pixel effects. Next, the median of each column's background rate is subtracted from each row to remove the $1/f$ noise for timescales longer than a row read time (5.24 ms). The correction was applied to each group so that $1/f$ noise would not be detected as spurious jumps or cosmic rays by the pipeline. We used all pixels more than 201 pixels from the source to estimate the background and $1/f$ noise, then subtracted the median of each row from all pixels in that row. Stage 2 of jwst was skipped, as it only changes the rates from ADU per second to physical units and conducts flat fielding.

[2] https://tshirt.readthedocs.io/en/latest/

[3] https://jwst-pipeline.readthedocs.io/en/latest/jwst/refpix/index.html

This does not affect the relative measurements of the light curve (due to the high pointing precision) and allows for comparison to detector-level effects.

For the photometric extraction, we used a source radius of 79 pixels and a background annulus of 79 to 100 pixels. We performed a 2D Gaussian fit to determine the centre of the aperture.

**Long-Wavelength Spectroscopy**

We performed three independent long-wavelength data reductions, using the `Eureka!`, `HANSOLO`, and `tshirt` pipelines.

The reference files in the Calibration Reference Data System (CRDS) at the time of our analysis included a linear solution for wavelength as a function of $x$ coordinate (the dispersion direction), but this is not strictly accurate at the blue end. For all methods, we use commissioning program 1076 to derive a third-degree polynomial wavelength solution that uses the Pfund and Bracket Hydrogen Series in the planetary nebula IRAS 05248-7007. The residuals in this solution are $\lesssim 0.1$ nm and the stellar absorption lines in WASP-39 agree with the solution to within 1 nm. The difference between the corrected wavelengths and the original wavelength solution is almost zero at the red end of the spectrum, but increases to about 50 nm at the blue end.

**Eureka! Long-Wavelength Reduction**

We investigated several variations of the `Eureka!` long-wavelength data reduction to minimise the MAD of the final extracted light curves, with different settings for cosmic-ray jump detection, identifying the spectral trace, the aperture size for spectral extraction, the region for background subtraction, and limits for outlier rejection. Here we present details of the data reduction that produced the spectrum shown in the main body of the paper.

Stages 1 and 2 were identical to the jwst pipeline, with the exception of increasing the rejection threshold during jump detection to 6σ. In Stage 3, we first trimmed the data to a subarray extending from pixels $4 - 64$ in the cross-dispersion direction and $4 - 1704$ in the spectral direction. We then masked any pixels with NaN values for the flux or error. We fit the spectral trace with a Gaussian profile and corrected for the curvature of the trace to the nearest integer pixel. We excluded a region 14 pixels wide on either side of the spectral trace from the background calculation and performed a column-by-column linear fit to subtract the background. We used a double-iteration 7σ threshold for outlier rejection of the sky background along the time axis during background subtraction. Additionally, we used a 7σ threshold for outlier rejection during the polynomial fit to the background. To obtain the spectrum, we constructed a normalised spatial profile using the median of all data frames, then used optimal extraction[53] on an aperture with a half-width of 9 pixels. For the optimal extraction, we rejected outliers above a 10σ threshold. Extended Data Figure 3 shows the curvature-corrected, background-subtracted, median frame with indicated background and aperture regions.

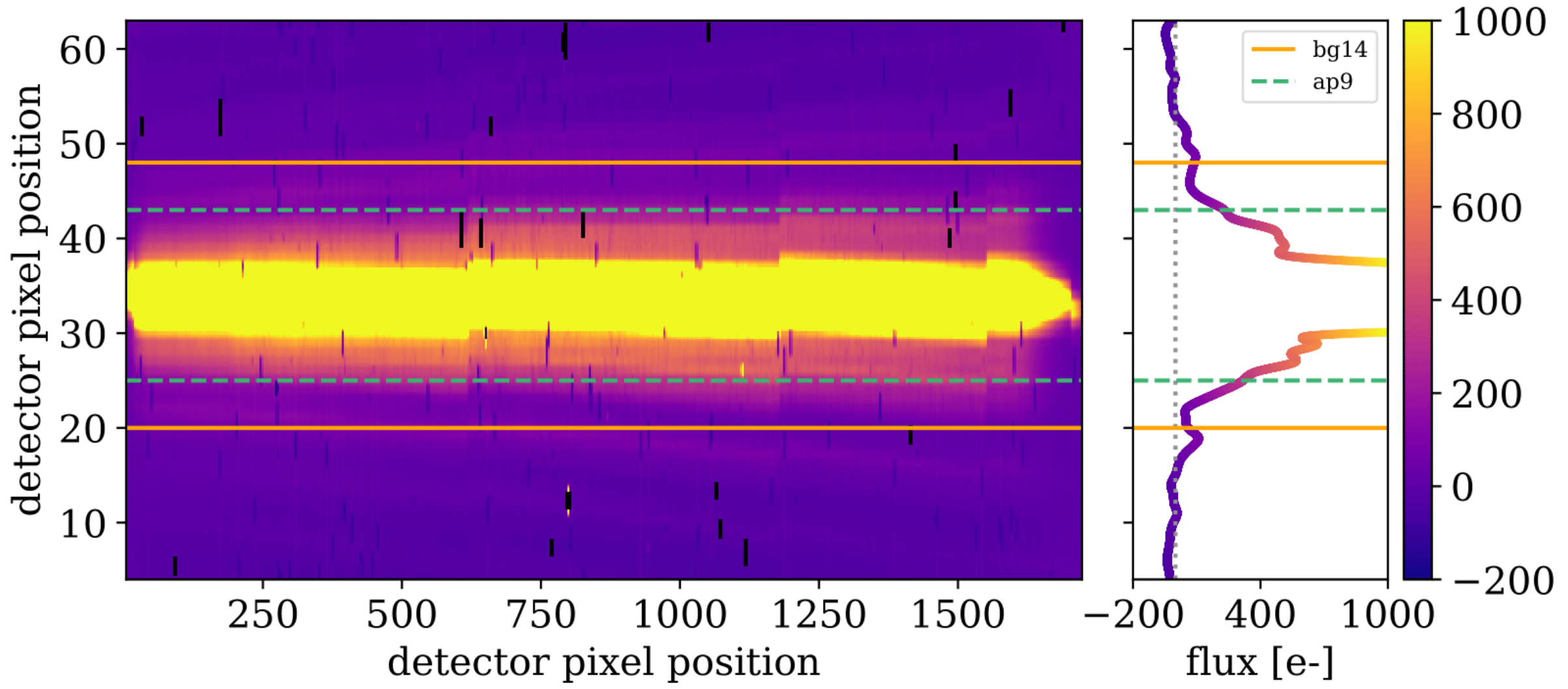

**Extended Data Figure 3: Curvature-corrected, background-subtracted, median frame.** We perform optimal spectral extraction on the pixels in between the green dashed lines. We use the pixels outside of the two orange solid lines for background subtraction. The flux spans -200 – 1000 electrons, thus drawing attention to the residual background features. **Right: Vertical slice depicting the flux averaged over detector pixels 855 to 865.** The background region clearly demonstrates some low-level residual structure.

### HANSOLO Long-Wavelength Reduction

The `HANSOLO` (atmospHeric trANsmission SpectrOscopy anaLysis cOde) pipeline was originally developed to analyse ground-based transmission spectra observed with 8m-class telescopes[54,55] and was adapted to enable its use on NIRCam data. `HANSOLO` begins with the calibrated rateints.fits outputs of jwst Stage 1.

We used the LACOSMIC algorithm[56] to remove cosmic ray effects from the two-dimensional images and identified the spectral trace using a Moffat function fit to each column. To remove the sky, we fitted and subtracted a linear trend from each column, excluding from the fit a region of 20 pixels on either side of the trace centre. We then extracted the spectrum by summing over an aperture with a half-width of 3 pixels. The spectra from different images were aligned with each other using cross correlation. To correct outlier pixels, each spectrum was normalised to account for the effect of the transit on the time series. Outliers $> 3\sigma$ away from the mean were removed from the time series of each wavelength point in the normalised spectra and replaced with the median value over time. We then rescaled the spectra to their original amplitudes.

### tshirt Long-Wavelength Reduction

As with the short-wavelength reduction, a few modifications were made to the Stage 1 CalWebb ramps-to-slopes pipeline. ROEBA subtraction reduced $1/f$ noise (described above for photometry); however, only pixels 1847 to 2044, which are on the rightmost amplifier, are available as low-illumination background.

For Stage 3, `tshirt` performed optimal spectral extraction weighted by the covariance between pixels[27]. We used a spectral aperture centred at pixel 34 in the spatial direction with a half-width of 5 pixels. We selected the background region to extend between pixels 5-24 and 44-65 in the spatial direction. The background was fit with a column-by-column linear trend with 3-sigma clipping. For the spectral extraction, we fit the spatial profile with a cubic spline with 20 knots and an outlier rejection threshold of 30σ. If a pixel was deemed an outlier either by the “DO_NOT_USE” data quality flag or by the spatial profile outlier detection, the rest of the spatial profile was weighted by the reference profile to ensure that the flux was conserved. For the covariance weighting, a correlation of 8% was assumed between pixels as measured by background pixels’ noise properties.

**Data Analysis and Fitting**

We used both `Eureka!` and `tshirt` to fit the short-wavelength light curves. In both cases, the light curves were fit with models that included both the transit and systematic noise. However, in order to investigate the effect of different systematic models on the resulting spectra, each fit used a slightly different noise model. Extended Data Table 1 summarises the systematics models which were used in each short-wavelength fit.

For the long-wavelength fits, we summed the data into 15 nm bins (~ 15 pixels). We experimented with bins as small as 10 nm, but found that reducing the bin size below 15 nm led to poor constraints on the limb darkening and added additional scatter to the resulting spectrum. Extended Data Figure 4 shows that the noise is primarily Gaussian out to long time scales of order the length of ingress/egress. Additionally, we created a white light curve by summing the extracted spectra over the entire 2.420 – 4.025 μm wavelength region. We experimented with different wavelength cut-offs but chose to extract spectra in this wavelength region because the low instrument throughput affected the quality of the extracted light curves beyond this region. Extended Data Figure 5 shows all reduced transmission spectra with one bin added on the blue end and two added on the red end, as well as the relative throughput at the wavelengths of these bins. This figure demonstrates the large error bars derived from data near the edges of the NIRCam/F322W2 bandpass. Therefore, we recommend that future works limit extracted spectra to the wavelength region between 2.420 – 4.025 μm.

**Extended Data Table 1:** Details of the two methods used to fit the short-wavelength photometry. Abbreviations for priors are as follows: U=uniform prior, with numbers indicating lower and upper limits; N=normal prior, with numbers indicating mean and sigma; LN=log-normal prior, with numbers indicating mean and sigma.

| | Fitting Method | |
|---|---|---|
| | `Eureka!` | `tshirt` |
| **Noise Parameters and Priors** | | |
| Polynomial in time ($c_0$,$c_1$,etc.; unitless, days$^{-1}$, etc.)* | 1$^{st}$-order $c_0$: U,0.9,1.1 $c_1$: U,-0.1,0.1 | 2$^{nd}$-order $c_0$: N,24,0.24 $c_1$: N,0,0.576 $c_2$: N,0,0.144 |
| **System Parameters and Priors** | | |
| Planet-to-star radius ratio ($R_p/R_s$, unitless) | U,0,0.3 | LN,$\ln(0.08)$,0.5 |
| Period ($P$, days) | fixed to 4.05527999[14] | N,4.05527999,$7\times10^{-7}$[14] |
| Mid-transit time ($T_c$, BJD$-2459783$) | U,0.45,0.55 | N,0.5005,0.0007 |
| Inclination ($i$, °) | U,80,90 | N,87.93,0.14[14] |
| Scaled semi-major axis ($a/R_s$, unitless) | U,2,20 | N11.55,0.13[14] |
| Limb darkening law used | Kipping 2-parameter[57] | Kipping 2-parameter[57] |
| Limb darkening parameters ($u_1$,$u_2$) | U,0,1 | Uninformative priors[57] |
| **Fit Results** | | |
| transit depth (ppm) | $21103 \pm 85$ | $21177 \pm 53$ |

* Subscripts 0, 1, etc. indicate the 0th, 1st, etc. order terms in polynomial models

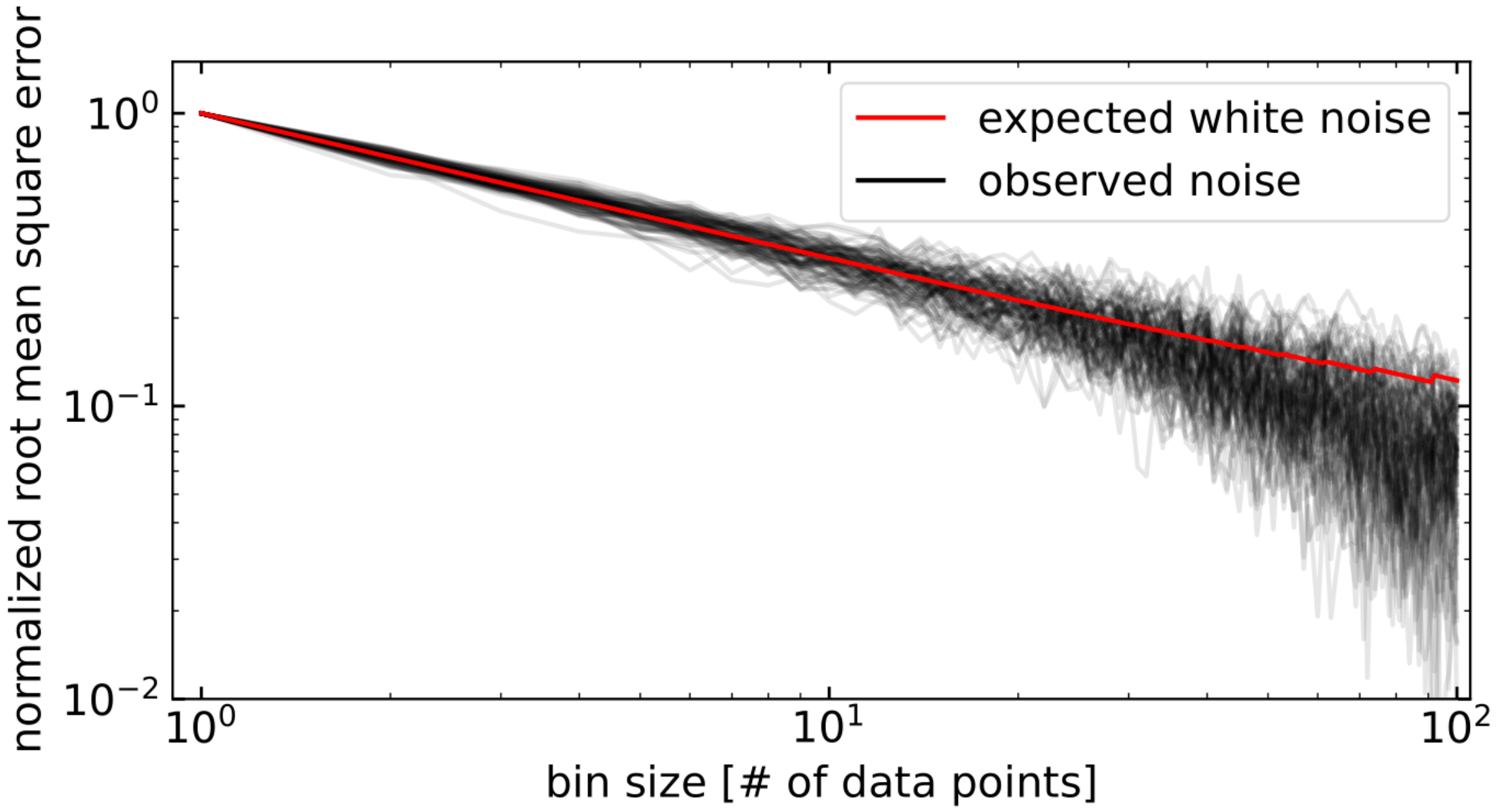

**Extended Data Figure 4: Normalised root mean square error as a function of bin size for all spectroscopic channels.** The red line shows the expected relationship for perfect Gaussian white noise. The black lines show the observed noise from each spectroscopic channel for the `Eureka!` long wavelength reduction. Values for all channels are normalized by dividing by the value for a bin size of 1 in order to compare bins with different noise levels. The black lines closely follow the red line out to large bin sizes of ≈30 (≈0.5-hr time

scales), which demonstrates that the residuals to the fit are dominated by white Gaussian noise.

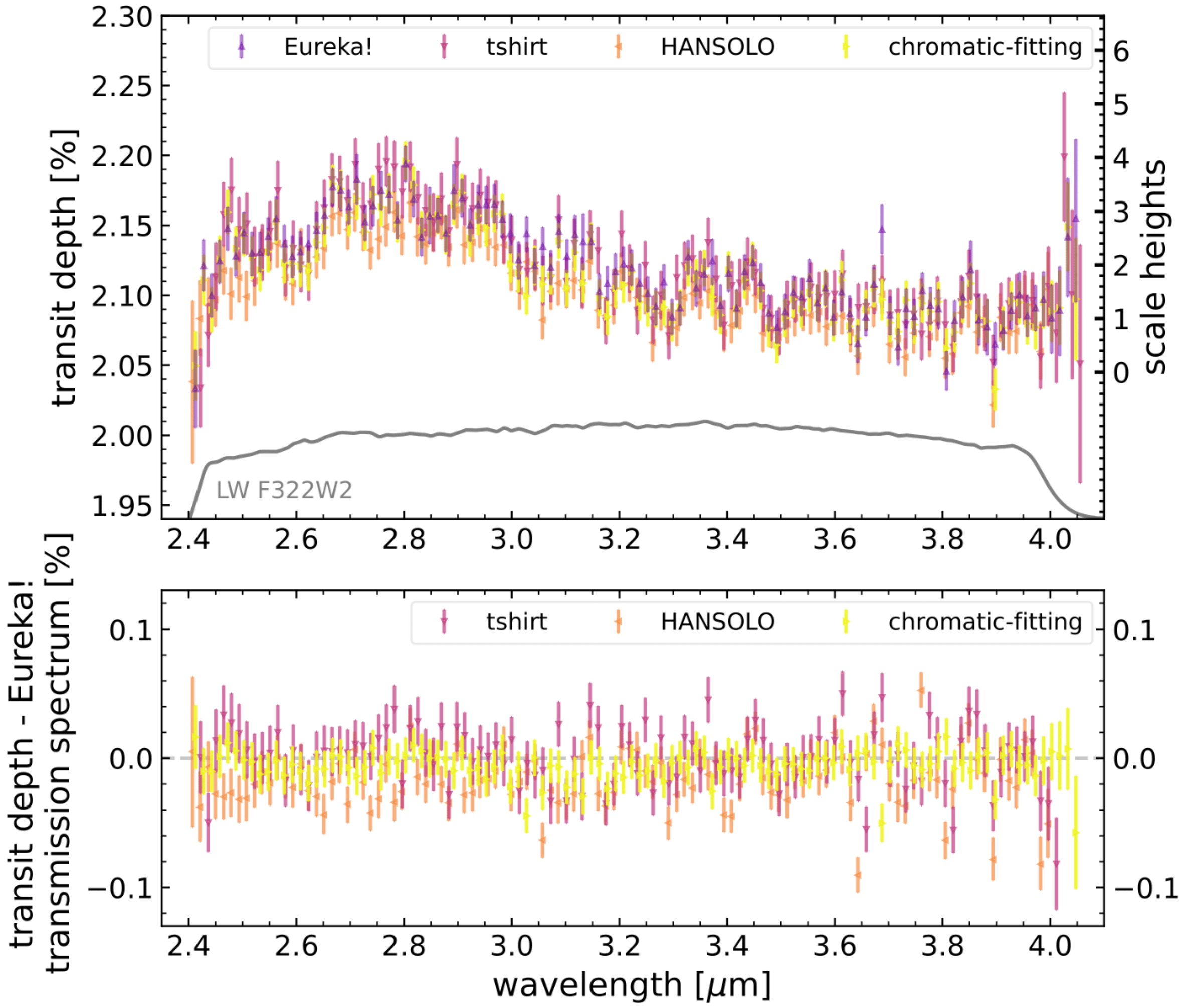

**Extended Data Figure 5:** Top panel: Transmission spectra from our reductions when including additional data on the blue and red edges (now spanning 2.405 – 4.055 μm), demonstrating the large error bars and diverging data points near the edges of the NIRCam bandpass in the LW spectroscopic channel. The differences in retrieved transmission spectra by subtracting the `Eureka!` spectrum from the other three reduced spectra shown in the top panel. This shows the strong agreement between the spectra; however, we do note minor disagreements at shorter wavelengths that we attribute to differences in the treatment of limb-darkening effects within the individual fitting methods.

We fit the long-wavelength light curves using four independent pipelines: `chromatic-fitting`, `Eureka!`, `HANSOLO`, and `tshirt`. `chromatic-fitting` is an open-source[4] Python tool to perform light-curve fitting, built on the data visualiser

[4] https://github.com/catrionamurray/chromatic_fitting/

`chromatic` (Berta-Thompson, in prep.[5]). For this work, `chromatic-fitting` light-curve fitting was applied to a `Eureka!` data reduction. As with the short-wavelength fits, we fit the long-wavelength light curves with models that include different noise parameterisations. Extended Data Table 2 summarises the systematics models that were used in each long-wavelength fit.

**Extended Data Table 2:** Details of the four fitting methods used to fit the long-wavelength spectroscopy. Abbreviations for priors are as follows: U = uniform prior, with numbers indicating lower and upper limits; LU = log-uniform prior, with numbers indicating lower and upper limits; N = normal prior, with numbers indicating mean and sigma; LN = log-normal prior, with numbers indicating mean and sigma. The notation ``spec-fixed" indicates that a value was fit in the white light curve and fixed to the best-fit value for the spectroscopic light curves.

| | Fitting Method | | | |
|---|---|---|---|---|
| | *chromatic-fitting* | Eureka! | HANSOLO | tshirt |
| **Noise Parameters and Priors*** | | | | |
| Polynomial in time ($c_0$,$c_1$,etc.; unitless, days$^{-1}$, etc.) † | 2nd-order $c_0$: N,1.0,0.01<br>$c_1$: N,0.0,1e-4 $c_2$: N,0.0,1e-4 | 1st-order $c_0$: N,1.0,0.001<br>$c_1$: N,0.0,0.01 | 0th-order $c_0$: U,0.8,1.2 | 2nd-order $c_0$: N,24,0.24<br>$c_1$: N,0,0.576 $c_2$: N,0,0.144 |
| Polynomial with drift in y position ($y_0$,$y_1$,etc.; unitless, pixels$^{-1}$, etc.) | 2nd-order $y_0$: fixed to 0.0<br>$y_1$: N,0,1e-4 $y_2$: N,0,1e-4 | N/A | N/A | N/A |
| Gaussian Process 3/2 Matern kernel model (amplitude $A$, correlation length $L$; unitless, days) | N/A | N/A | $A$: LU,$10^{-20}$,$10^{-5}$<br>$L$: LU,$10^{-10}$,1 ‡ | N/A |
| Multiplier to the expected noise level from Stage 3 ($s$, unitless) | N,1,0.1 | N,1.1,0.1 | N/A | N/A |
| **System Parameters and Priors** | | | | |
| Planet-to-star radius ratio ($R_p/R_s$, unitless) | N,0.145,0.05 | N,0.145,0.05 | U,0,1 | LN,$\ln(0.08)$,0.5 |
| Period ($P$, days) | fixed to 4.05527999[14] | fixed to 4.05527999[14] | fixed to 4.05527999[14] | N,4.05527999,$7\times10^{-7}$[14] |
| Mid-transit time ($T_c$, BJD$-2459783$) | N,0.5,0.02; spec-fixed | N,0.5,0.05; spec-fixed | U,0.45,0.55; spec-fixed | N,0.5005,0.0007 |
| Inclination ($i$, $^\circ$) | N/A | N,87.93,0.25; spec-fixed | N/A | N,87.93,0.14[14] |
| Scaled semi-major axis ($a/R_s$, unitless) | N/A | N,11.55,1; spec-fixed | N/A | N,11.55,0.13[14] |
| Impact parameter ($b$, unitless) | U,0,1.145; spec-fixed | N/A | U,0,1; spec-fixed | N/A |
| Limb darkening law used | Kipping 2-parameter[57] | quadratic | quadratic | Kipping 2-parameter[57] |
| Limb darkening parameters ($u_1$,$u_2$) | N, $\mu$ from ExoTiC-LD[63–65] model, 0.05 | U,-1,1; $u_1$ spec-fixed | N, $\mu$ from ExoTiC-LD[63–65], 0.1; $u_1$ spec-fixed | Uninformative priors[57] |
| **Fit Statistics** | | | | |
| Median error bar on final spectrum (ppm) | 121 | 135 | 137 | 180 |

* Note that different fitting methods used different parameterisations of the planetary orbit and noise model, so not all methods fit for all of the listed parameters. Parameters marked with ``N/A" were not fit in that method and were instead derived from the other fitted parameters.

[5] https://github.com/zkbt/chromatic/

† Subscripts 0, 1, etc. indicate the 0th, 1st, etc. order terms in polynomial models.
‡ The GP model was only applied to the white light curve. For the spectroscopic light curves, the divide-white method62 was used to remove the GP systematics.

For all fits, the parameters were estimated with a Markov Chain Monte Carlo (MCMC) fit, using either the `emcee` Python package[58] (for fits performed with `Eureka!`), the `pymc3` Python package[59] (implemented through the `Exoplanet` code[60,61], for fits performed with `chromatic-fitting` or `tshirt`), or the `CONAN` Python package[54,55] (for fits performed with `HANSOLO`). The number of free parameters and the resulting differential MADs of the light curves from each fit are also listed in Extended Data Tables 1 and 2. The best-fit parameters from the white light-curve fits are given in Extended Data Table 3.

**Extended Data Table 3:** Best-fit orbital parameters from both short-wavelength (SW) and white-light long-wavelength (LW) fits.

**Table 3 Best-fit orbital parameters from both short-wavelength (SW) and white-light long-wavelength (LW) fits.**

| Pipeline | $T_c - 2459783$ (BJD) | $R_p/R_s$ | $a/R_s$ | $i$ (°) |
|---|---|---|---|---|
| *Eureka!* SW | $0.50153 \pm 0.00003$ | $0.1453 \pm 0.0003$ | $11.43 \pm 0.05$ | $87.77 \pm 0.06$ |
| *tshirt* SW | $0.501540 \pm 0.000017$ | $0.14552 \pm 0.00018$ | $11.458 \pm 0.026$ | $87.79 \pm 0.03$ |
| *chromatic-fitting* LW | $0.501616 \pm 0.000024$ | $0.14531 \pm 0.00019$ | $11.43 \pm 0.20$ | $87.78 \pm 0.52$ |
| *Eureka!* LW | $0.501582 \pm 0.000032$ | $0.14588 \pm 0.00030$ | $11.381^{+0.055}_{-0.054}$ | $87.748^{+0.065}_{-0.063}$ |
| *HANSOLO* LW | $0.501624^{+0.000072}_{-0.000080}$ | $0.14482^{+0.00048}_{-0.00049}$ | $11.407^{+0.059}_{-0.061}$ | $87.802^{+0.071}_{-0.065}$ |
| *tshirt* LW | $0.501610 \pm 0.000014$ | $0.14563 \pm 0.00016$ | $11.44 \pm 0.02$ | $87.77 \pm 0.02$ |

In the process of performing the fits to the long-wavelength data, we regularly found that the best-fit transmission spectra were shifted vertically for different limb-darkening parameterisations and, for some reductions, exhibited changes in the apparent size of the water feature. In particular, we found that light-curve fits with all limb-darkening coefficients fixed to outputs from `ExoTiC-LD`[63–65] could result in a biased planet spectrum and might present a higher level of time-correlated noise in the residuals. We attribute this to a combination of *JWST*'s high-precision light curves and deficiencies in the stellar limb-darkening models to accurately represent WASP-39[66,67]. Therefore, the results presented here use the quadratic limb darkening law, in its classical form or reparameterised by Ref. [57], with one or both coefficients as free parameters. We confirmed that these parameterisations produce transmission spectra that are consistent both with each other and with the spectra resulting from using more complex limb darkening parameterisations, such as a four-parameter law with either fixed or free parameters[68]. We therefore recommend that future transmission spectrum analyses with NIRCam use similar methods. Limb-darkening conclusions from the full Transiting Exoplanet Community ERS program will be discussed further by Espinoza et al. (in prep.).

The final fitted light curves are shown in Extended Data Figure 6 and the final transmission spectra are shown in Figure 2. Both the short wavelength and long wavelength datasets are also available in our Reproducible Research Compendium on Zenodo at https://doi.10.5281/zenodo.7101283. The median difference between each transmission spectrum and the `Eureka!` spectrum is $0.87\sigma$ (using the maximum error at each point), which demonstrates a remarkable level of agreement. Additionally, the residuals showed no evidence for time-correlated noise, as shown in Extended Data Figure 5.

For ease of interpretation, we compared our atmospheric models to only one transmission spectrum. We selected the `Eureka!` spectrum, as it was on average nearest the median spectrum (the median transit depth at each bin).

## Atmospheric Forward Modelling

To interpret the long-wavelength data from NIRCAM/F322W2, we performed $\chi^2$ fits to the transmission spectra using three grids of radiative-convective equilibrium models: `ATMO`[69–71], `PHOENIX`[72–74], and `PICASO 3.0`[75,76]. All models used a common set of planetary parameters, but had differing opacity sources, cloud treatments, and grid points, described in detail below. Each model was binned to the resolution of the data to perform the $\chi^2$ fitting. We performed these three independent model grid fits to fully vet our inferences about the atmospheric metallicity and the presence of specific molecular features within the data.

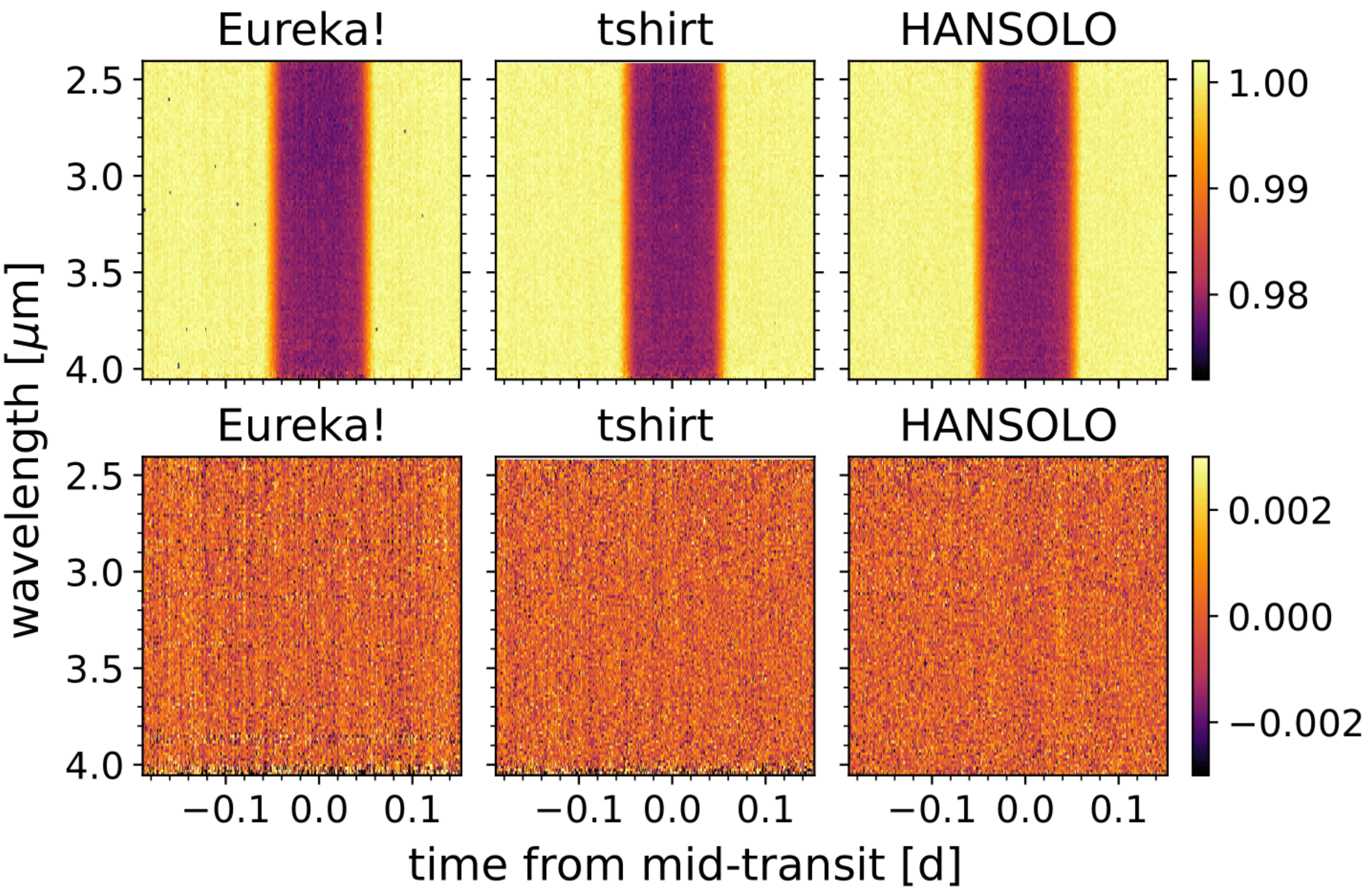

**Extended Data Figure 6:** Top Row: Time-series NIRCam data for the WASP-39b system, from three independent spectral extractions. Color represents relative brightness at each time and wavelength, normalized by the median stellar spectrum. Bottom Row: Resulting

residuals after fitting the time-series NIRCam data.

### The PICASO 3.0, Vulcan, & Virga Model Grid

Our primary atmospheric model grid is built from the open-source radiative-convective equilibrium code PICASO (Planetary Intensity Code for Atmospheric Spectroscopy Observations)[75], version 3.0[76], which was developed from the Fortran-based EGP[77–79]. We used PICASO 3.0 to generate one-dimensional temperature-pressure profiles in thermochemical equilibrium. The base PICASO 3.0 forward model grid computes atmospheric mixing ratios using variations of planetary intrinsic temperatures ($T_{int}$) of 100 K, 200 K, and 300 K; C/O ratios of 0.229, 0.458, 0.687, 0.916; and atmospheric solar metallicity values of 0.1, 0.316, 1.0, 3.162, 10.0, 31.623, 50.119, and 100× solar. The PICASO grid assumes full day-night energy redistribution. To compute model transmission spectra from the atmospheric profiles, we used opacities described by Ref.[79] (see in particular Extended Data Table 2); which sources $H_2O$ from Refs.[80,81], $CH_4$ from Refs.[82–84], $CO_2$ from Ref.[85], and $H_2S$ from Refs.[82,86,87].

We then used the one-dimensional CHON-based chemical kinetics code VULCAN[34] and the cloud modelling code Virga[88], which is the Python implementation of the Eddysed cloud code[89], to post-process disequilibrium chemistry from mixing and photochemical products as well as the effect of clouds. These additional post-processed grids also include vertically constant eddy diffusivities ($K_{zz}$) of $10^5 - 10^{11}$ cm$^2$/s in steps of 2 dex, and both clear and cloudy models. For the Vulcan disequilibrium runs, we only computed model grid points for a select subset of metallicity values (1, 10, 50, and 100× solar) and C/O ratios (0.229, 0.458, 0.687). We found that neither the cloudy nor clear disequilibrium grids from VULCAN offered an improvement in the $\chi^2_\nu$ value. Given the sparseness of these pre-computed disequilibrium grid models, we left rigorous quantification of self-consistent disequilibrium chemistry in the atmosphere of WASP-39b to future work.

Within PICASO, clouds are implemented both as grey absorbers and as Mie scatterers using temperature-relevant cloud condensate species from Virga. For the grey clouds, the grid specified a cloud optical depth ($\tau_{cloud}$) between 1 and 0.1 bar ranging from $\tau_{cloud} = 3.2\times 10^{-6}$ to 1 in steps of 0.1 dex across all wavelengths. For clouds of specific condensates, we used Virga to compute log-normal particle size distributions using sedimentation efficiency ($f_{sed}$) values of 0.6 to 10 for MnS, $Na_2S$, and $MgSiO_3$ along the range of $K_{zz}$. Smaller sedimentation efficiencies, $f_{sed}$, with larger eddy diffusivities, $K_{zz}$, generated more extended cloud decks and stronger cloud opacity.

### The PHOENIX Model Grid

We also used a grid of atmosphere models from the PHOENIX radiative-convective equilibrium code to fit the data[72–74]. Similar to the PICASO 3.0 grid, parameters including the day-night energy redistribution factors, interior temperature (200 K, 400 K), bulk atmospheric metallicity (0.1, 1, 10, 100× solar), and C/O ratio (136 grid points from 0.3 to 1) were varied. Aerosol properties were parameterized through a haze factor (0, 10× multi-gas Rayleigh scattering) and a grey cloud deck pressure level (0.3, 3, and 10 mbar). Models with molecular abundances quenched at 1 bar to simulate vertical mixing were also calculated. The grid also included rainout to account for species sequestered as condensates in the deep atmosphere. Opacities are described by Refs. [74,90] and taken from Ref. [86].

### the ATMO Model Grid

Similar to the model grids described above, we compared the data to a grid of models from the ATMO radiative-convective-thermochemical equilibrium code[69–71,91]. The ATMO grid used similar atmospheric and aerosol parameterisations to those used in the PHOENIX grid and also included rainout that accounts for species condensed in the deep atmosphere. Also included are day-night energy redistribution factors (0.25, 0.5, 0.75, and 1; with 1 as full redistribution), atmospheric metallicity (0.1, 1, 10, 100× solar), interior temperature (100, 200, 300, 400 K), C/O ratio (0.35, 0.55, 0.7, 0.75, 1.0, 1.5), cloud scattering factor (0, 0.5, 1, 5, 10, 30, 50× $H_2$ Rayleigh scattering at 350 nm between 1 and 50 mbar pressure levels), and a haze scattering factor (1, 10× multi-gas Rayleigh scattering). Opacities for $H_2O$, $CO_2$, and $CH_4$ are taken from Ref. [81–84] and $H_2S$ from Ref. [86].

## Grid Fits to *JWST*/NIRCam Data

We applied each of our three grids – ATMO, PHOENIX, and PICASO 3.0 – to fitting the NIRCam F322W2 spectrum (2.4 – 4.0 µm). In doing so, we found that the models strongly favoured a solar- or super-solar-metallicity atmosphere (1 – 100× solar), a sub-stellar C/O ratio (≤ 0.35), and substantial contribution from clouds, which are parameterized differently by each model grid (see each grid description above). We show the best fits from each model grid in Extended Data Figure 7. This interpretation is in agreement with the results using *JWST*'s NIRSpec/PRISM instrument from 3.0 – 5.0 µm[7], improving on the wider spread from previous *HST*-only[13–15,17,18,92] or *HST* and ground-based optical interpretations[16].

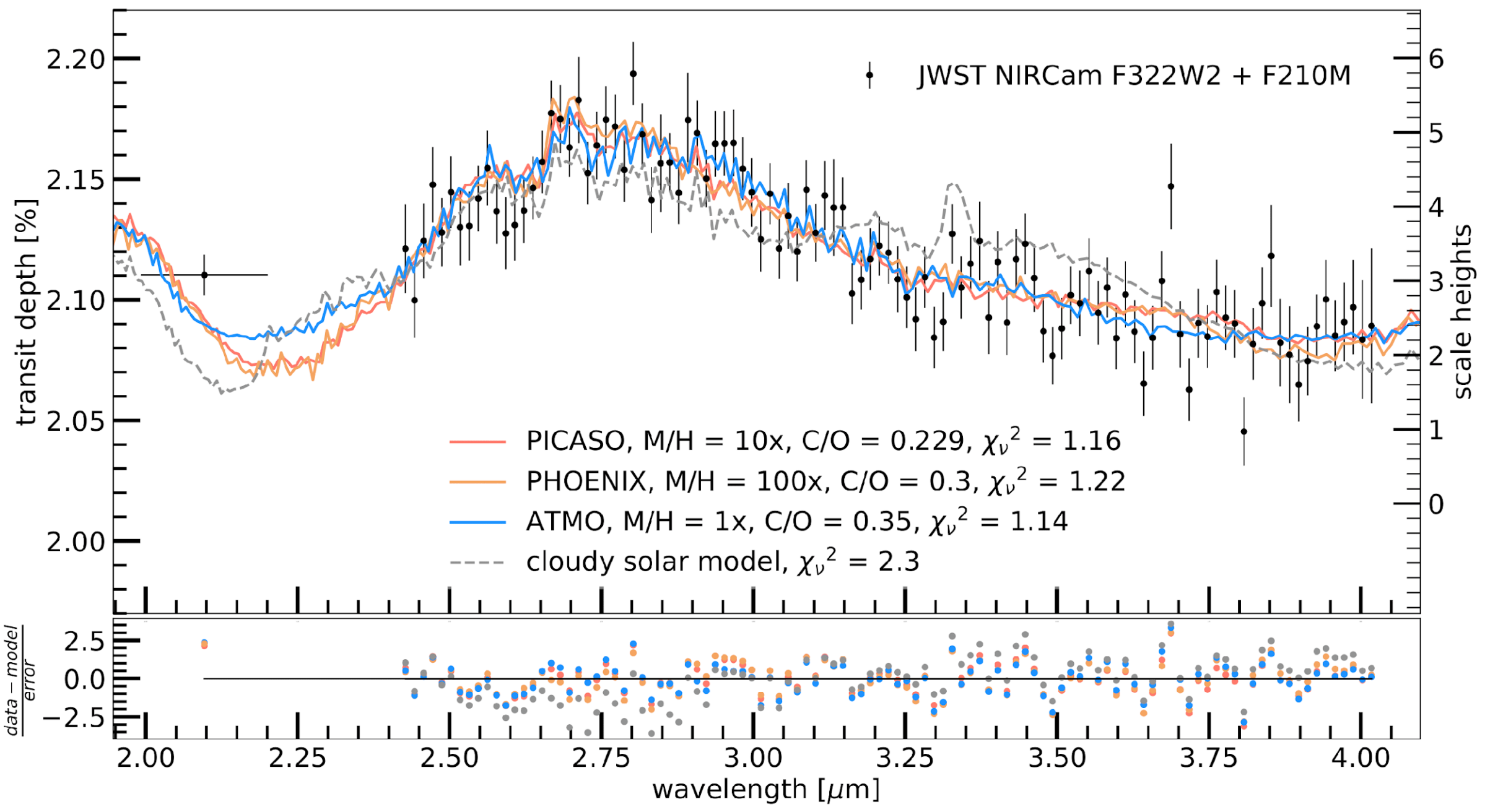

**Extended Data Figure 7: Measured transmission spectrum compared to atmospheric forward model grids.** Top: The single best fit for each model grid (shown as solid colored lines; `PICASO 3.0`, `ATMO`, `PHOENIX`), fits the planet spectrum (`Eureka!` reduction) with $\chi^2_\nu \leq 1.22$. All single best fits prefer at least solar metallicity and substantial cloud cover. Also shown as a grey dashed line is a solar metallicity, stellar C/O ratio atmospheric model, demonstrating the lack of methane absorption seen in the spectrum. Because we can put an upper limit on the $CH_4$ abundance, the preferred C/O ratio found by the model grids is sub-stellar. Bottom: Residuals of each best fit, shown as the model spectrum subtracted from the reduced spectrum and divided by the uncertainty in transit depth. The residuals show wavelength-dependent correlations, the origin of which are unknown and left for a future study.

For the NIRCam-only fit, the `PICASO` grey-cloud scheme produced a slightly better best fit ($\chi^2_\nu = 1.16$) than the `PICASO` + `Virga` more realistic clouds ($\chi^2_\nu = 1.23$), both of which were preferred to the clear-model best fit (100× solar) with $\chi^2_\nu = 1.53$. The `Virga` best-fit grid resulted in an atmosphere of 1× solar metallicity, C/O = 0.229, $f_{sed} = 0.6$, and $K_{zz} = 10^9$ cm/s$^2$. This `Virga` best-fit model consists of clouds of MnS and $MgSiO_3$ with deep (≥ 100 bars) cloud bases and diminishing optical depth up to ~ mbar pressures.

The best-fit equilibrium model from the `PHOENIX` grid had 100× solar metallicity, a C/O ratio of 0.3, and a cloud deck at 3 mbar. Cloudy models were generally preferred over clear models, but not with statistical significance ($\chi^2_\nu$ of 1.25 compared to 1.22). The `PHOENIX` grid finds best fits with very high metallicity (100× solar), so this low confidence regarding clouds reflects the cloud-metallicity degeneracy inherent in data restricted to narrow wavelengths[e.g., 93], as well as potentially the sparseness of the model grid.

For the `ATMO` grid, the best-fit equilibrium model to the NIRCam spectrum was 1× solar metallicity, a C/O ratio of 0.35, a cloud factor of 5 and a haze factor of 1. As with the other two grids, strongly cloudy models (cloud factor of ≥5) were preferred to clear models ($\chi^2_\nu$ of 1.1 vs 1.2).

**HST+NIRCam**

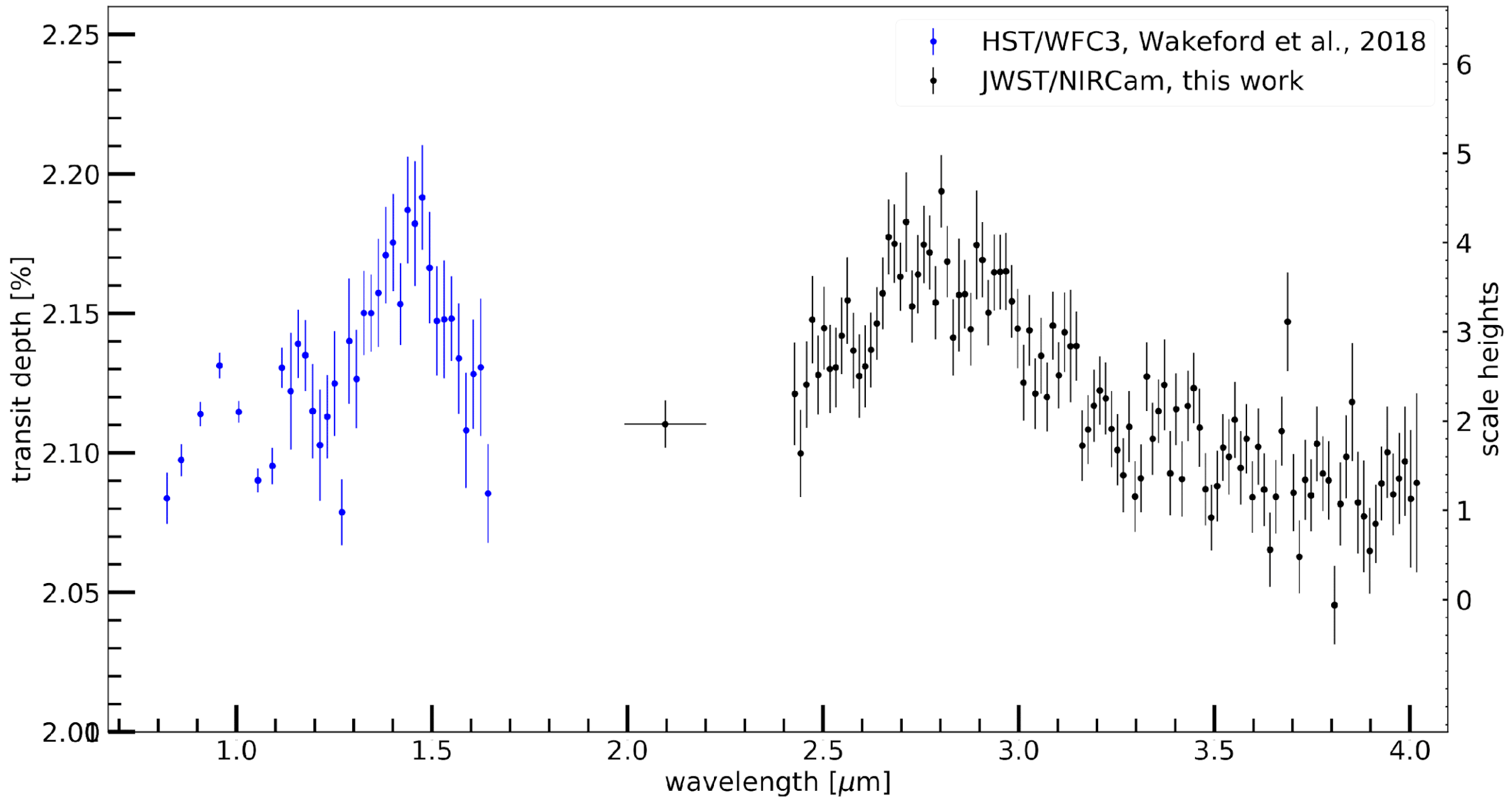

**Extended Data Figure 8: Our JWST/NIRCam spectrum compared to existing HST/WFC3 data.** As in Extended Data Figure 7, but with the addition of HST/WFC3 data from 0.8 to 1.65 μm, showing the comparable precision and complementary wavelength coverage offered by the combination of NIRCam and HST/WFC3.

In Extended Data Figure 8, we show the comparison between the spectra of HST/WFC3 (G141 and G102, covering 0.8 – 1.65 μm) and JWST/NIRCam (F210M+F322W2, 2.0 – 4.0 μm). We chose to only show Wide Field Camera 3 (WFC3) observations from *HST*, as these are of higher precision than observations from the Space Telescope Imaging Spectrograph (STIS) or ground-based data[13]. Additionally, as *HST*/WFC3 has the most archival exoplanet data of any instrument on *Hubble*, future JWST exoplanet programs will primarily rely on this *HST* instrument for inter-telescope comparisons or extending the wavelength coverage of *JWST* data. For example, the addition of optical and shorter wavelength near-infrared data can help break metallicity degeneracies by better constraining the presence and extent of clouds[13,e.g., 93]. High altitude clouds or hazes can be inferred from their particle sizes, where small particles scatter shorter wavelengths more efficiently e.g., [94,95], thus enabling the disentanglement of a very cloudy, low metallicity atmosphere from a less cloudy, high metallicity atmosphere[17].

## Molecular Detections

Once we found the "single best fit" for the `PICASO` grid to the NIRCam spectrum (10× solar, C/O = 0.229, grey cloud optical depth = $2.6\times10^{-3}$ from 1 to 0.1 bar), we used this as a base model to explore the significance of specific molecular detections. First, we tested whether we could improve the best fit in the presence or absence of $H_2O$, $CO_2$, $CH_4$, or $H_2S$. We reran the best-fit base model by zeroing out each of these species in turn, shown in Figure 3, and then repeating our $\chi^2$ analysis.

We found that while the presence of $H_2O$, $H_2S$, and $CH_4$ resulted in a better $\chi^2_\nu$ value, only $H_2O$ and $H_2S$ did so in a statistically meaningful way. Since $H_2S$ does not contain strong molecular features within the NIRCam wavelength range, the Gaussian residual fitting we perform for the detection significance of other molecules is not applicable, and we left its further quantification to more rigorous atmospheric retrieval analyses. Increasing the $CH_4$ abundance beyond that of the best-fit model also improved the $\chi^2_\nu$, though again not to high statistical significance.

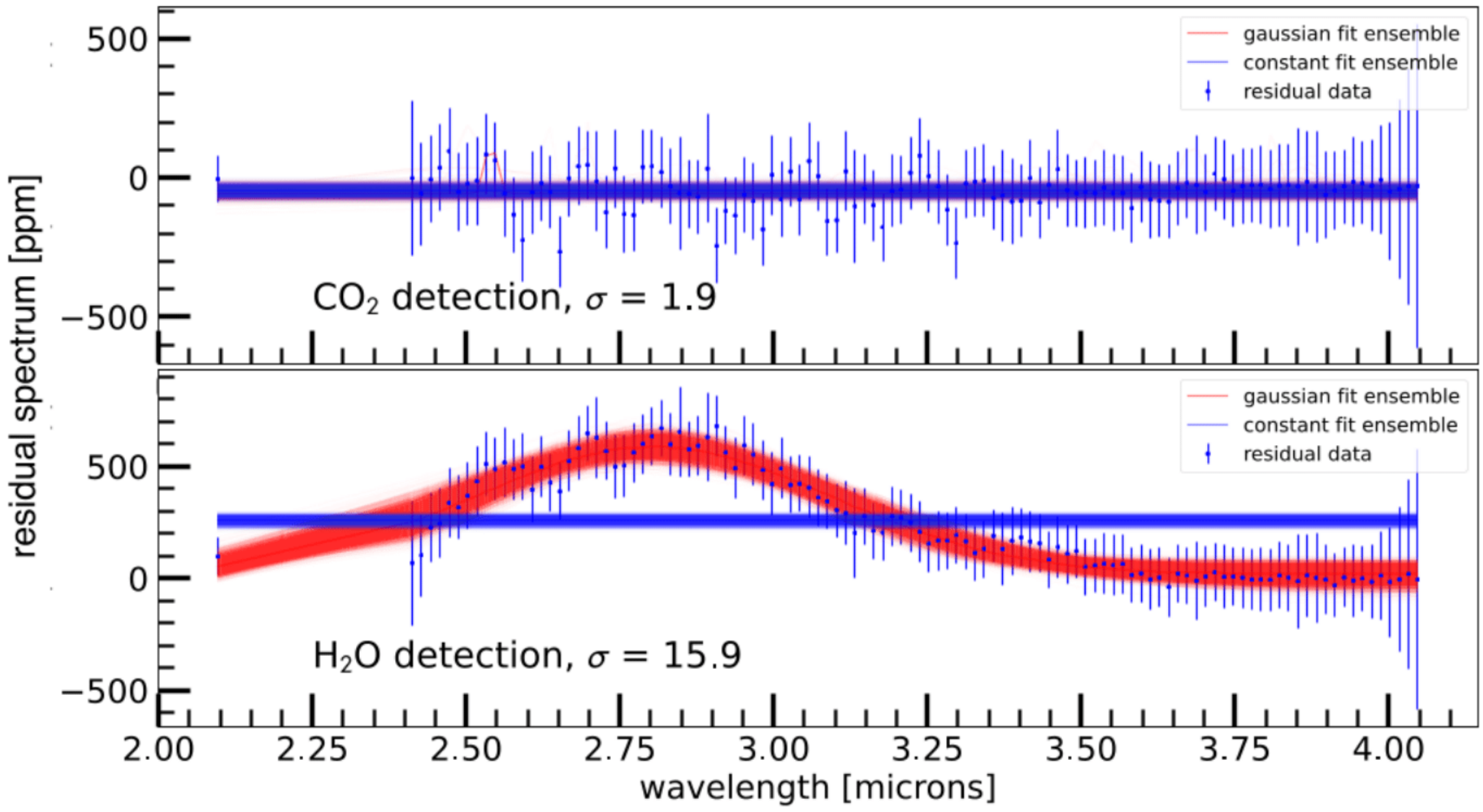

**Extended Data Figure 9: Gaussian residual fitting of $H_2O$ and $CO_2$.** The blue points show the residual features left after subtracting out the gas in question ($CO_2$, top, and $H_2O$, bottom) from the single best-fit model. The Gaussian model ensemble fit to the residual is shown in red; the best-fit Gaussian ensemble to a flat-line model is shown in blue. We strongly detect $H_2O$ at nearly 16σ and show weak evidence for $CO_2$ (small feature at 2.6 μm) at 1.9σ.

With the best fit in hand, we investigated the presence of individual molecular species. For molecular detection significances, we performed the same Gaussian residual fitting, shown in Extended Data Figure 9, as for the detection of $CO_2$ in the NIRSpec/PRISM 3.0 - 5.0 μm analysis[7]. We find a Bayes factor, ln($B$), of 123.2 between the Gaussian residual and constant models for $H_2O$ over the whole NIRCam wavelength range, corresponding to 15.9σ, a strong detection. For $CO_2$ we find ln($B$) of 0.82 between the Gaussian residual and constant models between 2.4 and 2.9 μm, or 1.9σ, which is a weak or non-detection[96]. $CO_2$ is strongly detected at 4.3 μm in the NIRSpec data for WASP-39b[7, Rustamkulov et al., submitted, Alderson et al., submitted], but the strong overlapping $H_2O$ band at 2.8 μm prevents NIRCam from making a significant $CO_2$ detection. Given our upper limit on $CH_4$ abundance, we also performed the same Gaussian residual fitting for $CH_4$ and find a weak or non-detection at approximately 2σ.

Both WASP-39b NIRSpec datasets[7, Rustamkulov et al., submitted, Alderson et al., submitted] observed evidence for a molecular feature near 4.0 μm, which is currently best explained by $SO_2$. The reddest data points (>4.025 μm) from NIRCam also show an increase that is consistent with this feature seen in the NIRSpec data. However, as shown in Extended Data Figure 5, these NIRCam data points have very large error bars because the detector throughput drops off dramatically past 4.0 μm. Future investigations to thoroughly explore the physicochemical likelihood of $SO_2$ in the atmosphere of WASP-39b must rely on wavelengths that can fully capture the complete absorption feature, which is beyond the reach of high fidelity NIRCam/F322W2 measurements.

**Data Availability**

The data used in this paper are associated with JWST program ERS 1366 (observation #2) and are available from the Mikulski Archive for Space Telescopes (https://mast.stsci.edu). We used calibration data from program 1076. All the data and models presented in this publication can be found at https://doi.10.5281/zenodo.7101283.

**Code Availability**

The codes used in this publication to extract, reduce and analyse the data are as follows: Batman[97], emcee[58], Eureka![52], jwst[98], chromatic, chromatic-fitting, PyMC3[59], Exoplanet[60,61], gCMCRT[99], CONAN[54,55], ExoTiC-LD[63–65], LACOSMIC[56], PICASO[75,76], Virga[88], VULCAN[34].

**Methods References**

**Acknowledgments:** This work is based in part on observations made with the NASA/ESA/CSA James Webb Space Telescope. The data were obtained from the Mikulski Archive for Space Telescopes (MAST) at the Space Telescope Science Institute (STScI), which is operated by the Association of Universities for Research in Astronomy, Inc., under NASA contract NAS 5-03127 for JWST. These observations are associated with program #1366. Support for this program was provided by NASA through a grant from STScI. This work is based in part on data collected under the NGTS project at the European Southern Observatory's La Silla Paranal Observatory. The NGTS facility is operated by a consortium of institutes with support from the UK Science and Technology Facilities Council (STFC) under projects ST/M001962/1, ST/S002642/1 and ST/W003163/1. This paper includes data collected by the *TESS* mission, obtained from MAST at STScI. Funding for the *TESS* mission is provided by the NASA's Science Mission Directorate. This article is supported by the UK Research and Innovation (UKRI) to fund open access through the University of Warwick.

**Author Contributions**

All authors played a significant role in one or more of the following: development of the original proposal, management of the project, definition of the target list and observation plan, analysis of the data, theoretical modelling, and preparation of this manuscript. Some specific contributions are listed as follows. NMB, JLB, and KBS provided overall program leadership and management. EA and KBS led the efforts for this manuscript. DS, EK, HW, IC, JLB, KBS, LK, MLM, MRL, NMB, VP, and ZBT made significant contributions to the design of the program. KBS generated the observing plan with input from the team. EvSc, NE, and TGB provided instrument expertise. BB, EK, HW, IC, JLB, LK, MLM, MRL, NMB, and ZBT led or co-led working groups and/or contributed to significant strategic planning efforts like the design and implementation of the pre-launch Data Challenges. AC, DS, EvSc, NE, NG, TGB, VP generated simulated data for pre-launch testing of methods. CAM, DP, EA, EvSc, JB, KBS, MD, MM, NN, SZ reduced the data, modelled the light curves, and produced the planetary spectrum. JG, JL, KO, NB, SEM, SM generated theoretical model grids for comparison with data. CAM, DP, EA, EvSc, JL, KBS, MM, NN, PJW, SEM made significant contributions to the writing of this manuscript. CAM, EA, KBS, MB, MM, PJW, SG, SEM, and ZBT generated figures for this manuscript. BVR, EL, GM, JB, SZ, ZBT contributed to the writing of this manuscript. JH, ML, NL, PG, PJW, RH, TD, TGB provided significant feedback to the manuscript.

**Competing Interests**

The authors declare no competing interests.

**Author affiliatons**

[1] Centre for Exoplanets and Habitability, University of Warwick, Coventry, UK
[2] Department of Physics, University of Warwick, Coventry, UK
[3] Johns Hopkins APL, Laurel, MD, USA
[4] Steward Observatory, University of Arizona, Tucson, AZ, USA
[5] NHFP Sagan Fellow
[6] Lunar and Planetary Laboratory, University of Arizona, Tucson, AZ, USA.
[7] Department of Physics & Astronomy, University of Kansas, Lawrence, KS, USA
[8] Instituto de Astrofísica de Canarias (IAC), Tenerife, Spain
[9] Departamento de Astrofísica, Universidad de La Laguna (ULL), Tenerife, Spain
[10] INAF- Palermo Astronomical Observatory, Piazza del Parlamento, Palermo, Italy
[11] Department of Astrophysical and Planetary Sciences, University of Colorado, Boulder, CO, USA
[12] Space Telescope Science Institute, Baltimore, MD, USA
[13] Département d'Astronomie, Université de Genève, Sauverny, Switzerland
[14] Max Planck Institute for Astronomy, Heidelberg, Germany
[15] Leiden Observatory, University of Leiden, Leiden, The Netherlands
[16] NASA Ames Research Center, Moffett Field, CA, USA
[17] Jet Propulsion Laboratory, Pasadena, CA, US

[18] School of Earth and Planetary Sciences (SEPS), National Institute of Science Education and Research (NISER), HBNI, Odisha, India
[19] Department of Physics, Utah Valley University, Orem, UT, USA
[20] Department of Astronomy & Astrophysics, University of California, Santa Cruz, Santa Cruz, CA, USA
[21] Astrobiology Program, UC Santa Cruz, Santa Cruz, CA, USA
[22] Department of Astronomy & Astrophysics, University of Chicago, Chicago, IL, USA
[23] Department of Astronomy, University of Wisconsin-Madison, Madison, WI USA
[24] Department of Physics and Institute for Research on Exoplanets, Université de Montréal, Montreal, QC, Canada
[25] Space Research Institute, Austrian Academy of Sciences, Graz, Austria
[26] INAF – Osservatorio Astrofisico di Torino, Pino Torinese, Italy
[27] Department of Astrophysical Sciences, Princeton University, Princeton, NJ, USA
[28] LSSTC Catalyst Fellow
[29] Department of Physics & Astronomy, Johns Hopkins University, Baltimore, MD, USA
[30] Earth and Planets Laboratory, Carnegie Institution for Science, Washington, DC, USA
[31] School of Physics, Trinity College Dublin, Dublin, Ireland
[32] Planetary Sciences Group, Department of Physics and Florida Space Institute, University of Central Florida, Orlando, Florida, USA
[33] Astrophysics Section, Jet Propulsion Laboratory, California Institute of Technology, Pasadena, CA, USA
[34] Division of Geological and Planetary Sciences, California Institute of Technology, Pasadena, CA, USA
[35] Department of Astronomy and Carl Sagan Institute, Cornell University, Ithaca, NY, USA
[36] School of Earth and Space Exploration, Arizona State University, Tempe, AZ, USA
[37] Center for Astrophysics | Harvard & Smithsonian, Cambridge, MA, USA
[38] Atmospheric, Oceanic and Planetary Physics, Department of Physics, University of Oxford, Oxford, UK
[39] Université Côte d'Azur, Observatoire de la Côte d'Azur, CNRS, Laboratoire Lagrange, France
[40] Department of Earth and Planetary Sciences, Johns Hopkins University, Baltimore, MD, USA
[41] School of Physics, University of Bristol, Bristol, UK
[42] NSF Graduate Research Fellow
[43] School of Physical Sciences, The Open University, Milton Keynes, UK
[44] BAER Institute, NASA Ames Research Center, Moffet Field, CA, USA
[45] Department of Physics, New York University Abu Dhabi, Abu Dhabi, UAE
[46] Center for Astro, Particle and Planetary Physics (CAP3), New York University Abu Dhabi, Abu Dhabi, UAE
[47] Mullard Space Science Laboratory, University College London, Holmbury St Mary, Dorking, Surrey, RH5 6NT, UK
[48] School of Physics and Astronomy, University of Leicester, Leicester
[49] European Space Agency, Space Telescope Science Institute, Baltimore, MD, USA
[50] Department of Physics and Astronomy, University College London, United Kingdom

[51] Centre for Exoplanet Science, University of St Andrews, St Andrews, UK
[52] Leiden Observatory, Leiden University, Leiden, The Netherlands
[53] Institute of Astronomy, Department of Physics and Astronomy, KU Leuven, Leuven, Belgium
[54] Anton Pannekoek Institute for Astronomy, University of Amsterdam, Amsterdam, The Netherlands
[55] Department of Physics and Astronomy, University of Delaware, Newark, DE, USA
[56] University Observatory Munich, Ludwig Maximilian University, Munich, Germany
[57] ARTORG Center for Biomedical Engineering, University of Bern, Bern, Switzerland
[58] Institute for Astrophysics, University of Vienna, Vienna, Austria
[59] Department of Astronomy, University of Maryland, College Park, MD, USA
[60] Department of Physics, Imperial College London, London, UK
[61] Imperial College Research Fellow
[62] Université Paris-Saclay, Université Paris Cité, CEA, CNRS, AIM, Gif-sur-Yvette, France
[63] Laboratoire d'Astrophysique de Bordeaux, Université de Bordeaux, Pessac, France
[64] Department of Astronomy, University of Michigan, Ann Arbor, MI, USA
[65] Department of Physics, University of Rome "Tor Vergata", Rome, Italy
[66] INAF - Turin Astrophysical Observatory, Pino Torinese, Italy
[67] Department of Physics and Astronomy, University of Exeter, Exeter, Devon, United Kingdom.
[68] SRON Netherlands Institute for Space Research, Leiden, the Netherlands
[69] Exzellenzcluster Origins, Garching, Germany
[70] Department of Earth, Atmospheric and Planetary Sciences, Massachusetts Institute of Technology, Cambridge, MA, USA
[71] Kavli Institute for Astrophysics and Space Research, Massachusetts Institute of Technology, Cambridge, MA, USA
[72] 51 Pegasi b Fellow
[73] Astronomy Department and Van Vleck Observatory, Wesleyan University, Middletown, CT, USA
[74] Institute of Astronomy, University of Cambridge, Cambridge, UK
[75] Maison de la Simulation, CEA, CNRS, Univ. Paris-Sud, UVSQ, Université Paris-Saclay, Gif-sur-Yvette, France
[76] Department of Physics, Brown University, Providence, RI, USA
[77] Planetary Science Institute, Tucson, AZ, USA
[78] Université de Paris Cité and Univ Paris Est Creteil, CNRS, LISA, Paris, France
[79] Department of Earth and Planetary Sciences, University of California Santa Cruz, Santa Cruz, California, USA